\documentclass[final,3p,times]{elsarticle}

\usepackage{amsfonts}
\usepackage{amsmath}
\usepackage{amssymb}
\usepackage{amsthm}
\usepackage{caption}
\usepackage{color}
\usepackage{epsfig}
\usepackage{enumerate}
\usepackage{enumitem}
\usepackage{float}
\usepackage{graphicx}
\usepackage{subcaption}
\usepackage{lineno}
\usepackage{multirow}
\usepackage{natbib} 
\usepackage{placeins}
\usepackage{setspace}
\usepackage{stmaryrd}
\usepackage{url}
\usepackage{verbatim}
\usepackage{algorithm}
\usepackage{hyperref}
\usepackage{algpseudocode}
\usepackage[table]{xcolor}
\usepackage{booktabs}
\usepackage{tikz}
\usepackage{smartdiagram}
\usesmartdiagramlibrary{additions}
\usetikzlibrary{positioning,shapes,arrows}
\usepackage{tikz}
\usetikzlibrary{automata}
\usesmartdiagramlibrary{additions}

\newtheorem{definition}{Definition}

\newcommand{\abs}[1]{\left\vert#1\right\vert}

\newcommand{\set}[1]{\left\{#1\right\}}

\newcommand{\pspace}{\Lambda}
\newcommand{\qspace}{\mathcal{Q}}
\newcommand{\pmeas}{\mu_{\pspace}}
\newcommand{\qmeas}{\mu_{\qspace}}

\newcommand{\preddens}{\pi_{\text{pred}}}
\newcommand{\obsdens}{\pi_{\text{obs}}}
\newcommand{\initdens}{\pi_{\text{init}}}
\newcommand{\updens}{\pi_{\text{update}}}

\newcommand{\predmeas}{\mathbb{P}_{\text{pred}}}
\newcommand{\obsmeas}{\mathbb{P}_{\text{obs}}}
\newcommand{\initmeas}{\mathbb{P}_{\text{init}}}

\newcommand{\qpca}{Q_{\mathrm{PCA}}}

\newcommand{\kl}{\mathrm{KL}_{\mathrm{DCI}}}
\newcommand{\er}{\mathbb{E_{\mathrm{init}}}(r(Q(\lambda))}
\newcommand{\epspr}{\epsilon_{\text{pred}}}
\newcommand{\epss}{\epsilon_{\text{samples}}}
\newcommand{\epskl}{\epsilon_{\text{KL}}}

\journal{TBD}

\begin{document}

\begin{frontmatter}

\title{Sequential Maximal Updated Density Parameter Estimation for Dynamical Systems with Parameter Drift}

\author[oden,tacc]{Carlos del-Castillo-Negrete}
\ead{carlosd@tacc.utexas.edu}
\author[oden]{Rylan Spence}
\ead{rylan.spence@utexas.edu}
\author[cud]{Troy Butler}
\ead{troy.butler@ucdenver.edu}
\author[oden]{Clint Dawson}
\ead{clint.dawson@utexas.edu}
\affiliation[oden]{organization={Oden Institute for Computational Engineering and Sciences, University of Texas at Austin},
            city={Austin},
            postcode={78712},
            state={TX},
            country={USA}}
\affiliation[tacc]{organization={Texas Advanced Computing Center},
            city={Austin},
            postcode={78758},
            state={TX},
            country={USA}}
\affiliation[cud]{organization={Department of Mathematical and Statistical Sciences, University of Colorado Denver},
            city={Denver},
            postcode={80202},
            state={CO},
            country={USA}}



\begin{abstract}
We present a novel method for generating sequential parameter estimates and quantifying epistemic uncertainty in dynamical systems within a data-consistent (DC) framework.
The DC framework differs from traditional Bayesian approaches due to the incorporation of the push-forward of an initial density, which performs selective regularization in parameter directions not informed by the data in the resulting updated density. 
This extends a previous study that included the linear Gaussian theory within the DC framework and introduced the maximal updated density (MUD) estimate as an alternative to both least squares and maximum a posterior (MAP) estimates.
In this work, we introduce algorithms for operational settings of MUD estimation in real- or near-real time where spatio-temporal datasets arrive in packets to provide updated estimates of parameters and identify potential parameter drift. 
Computational diagnostics within the DC framework prove critical for evaluating (1) the quality of the DC update and MUD estimate and (2) the detection of parameter value drift.
The algorithms are applied to estimate (1) wind drag parameters in a high-fidelity storm surge model, (2) thermal diffusivity field for a heat conductivity problem, and (3) changing infection and incubation rates of an epidemiological model.  
\end{abstract}



\end{frontmatter}



\section{Introduction}\label{sec:intro}

The impact of computational models for solving scientific and engineering challenges governed by principles of mechanics is tempered by the ability of these models to fit simulated predictions to observational data and quantify uncertainty in these predictions.
The ability of a model to fit observational data for a physical system is complicated by the fact that system behavior is often governed by key characteristics that have to be parameterized within the model and are hidden from direct observation.
The impact of perturbing the parameters in these models is usually observed indirectly via the simulation of model state data that can be post-processed into a set of Quantities of Interest (QoI) exhibiting sensitivities to these parameters.
This necessitates the formulation and solution of an inverse problem using discrepancies between (noisy) observational and simulated QoI data to quantify uncertainties in the model parameters and produce estimates of the parameters that best explain the observed data.

There are several complicating factors hindering the formulation and solution of inverse problems to provide updated estimates of parameters in operational settings.
First, in a real- or near-real time setting, it is usually the case that not all data are available simultaneously.
For instance, networks of sensors monitoring a dynamical system are often designed to transmit packets of data at specified time intervals to manage both data bandwidth limitations as well as ensure efficient power management of remote sensors. 
Second, different packets of data may exhibit unique sensitivities to parameters that require individualized post-processing into distinct QoI of potentially different dimensions.
This occurs, for instance, when external forces/conditions evolve over time that result in system states becoming more or less sensitive to certain characteristics appearing as model parameters.
This is also related to the third issue, which is that parameters may drift in time resulting in both qualitative and quantitative disruptions to the trajectories of state variables. 
Consequently, in order to be reliable in an operational setting, any method for monitoring and updating parameter estimates should also include mathematically justified quantitative diagnostics to assess the reliability of assumptions and computations in the inverse problem utilizing a given packet of data. 
This brings us to the two main contributions of this work designed to tackle the above complicating factors.

\begin{description}

    \item[Contribution 1:] We design an algorithm based on the data-consistent (DC) framework for quantifying uncertainties that permits for sequential parameter estimation in dynamical systems using the Maximal Update Density estimates originally presented in \cite{pilosov2023parameter}. 

    
    \item[Contribution 2:] Quantitative diagnostics based on both measure-theoretic and statistical principles are utilized to detect potential drift in parameter values driving dynamical systems as well as evaluate the suitability and reliability of the learned QoI and assumptions within the DC framework. 

\end{description}

The rest of this paper is organized as follows:
\begin{description}
    \item[Section~\ref{sec:background}:] A brief literature review of DC methods is provided. 
    The focus is on the Maximal Updated Density (MUD) method for estimating parameters in problems involving epistemic uncertainty that motivates the algorithms presented here.
    A definition of the parameter drift problem in terms of change point identification is also provided.

    \item[Section~\ref{sec:seq-estimation}:] 
    The foundational algorithms for DC inversion and MUD estimation are presented.
    The sequential MUD estimation algorithm follows after the computational features of the DC solution are discussed and utilized to develop a set of diagnostics that allow for a more comprehensive evaluation of solution quality and detection of parameter drift. 


    \item[Section~\ref{sec:results}:] Three practical examples are presented to demonstrate the flexibility of the proposed method to produce sequential estimates of parameter values for different dynamical systems. 
    These examples collectively demonstrate several features of the sequential algorithm and the DC framework.
    Each example individually demonstrates the proposed method's ability to reliably estimate parameter values that can produce observed data while simultaneously reducing the variance in these estimates as either more data are incorporated at each iteration or more iterations are utilized.
    We summarize a few high-level details of these examples here for ease of reference.

        \begin{enumerate}
            \item Storm Surge and Wind Drag - The first example presents an application of the algorithm to the problem of estimating wind-drag parameters for a model that best matches observed time-series of water elevation values. 
            In this example, we utilize the state-of-the-art ADvanced CIRCulation (ADCIRC) \cite{Luettich:1992,Westerink:1992} computational model for storm surge modeling that requires High Performance Computing (HPC) resources. 
            \item Thermal Diffusion - Here, the uncertain parameter is given by a random field, the thermal diffusivity $k(\boldsymbol{x})$, appearing in a standard variation of the heat equation. The thermal diffusivity is modeled via a Karhunen-Lo\`eve expansion \cite{huang2001convergence} to produce a nominally high-dimensional parameter space and demonstrate the efficacy of the sequential estimation algorithm on such spaces by utilizing a key feature of DC-based inversion to reduce the problem into a sequence of more computationally tractable lower-dimensional problems.
            \item SEIRS and Infectivity Rates - We present a compartmental epidemiological model that exhibits parameter drift to highlight the application of the quantitative diagnostics utilized to detect such drifts. We consider a scenario where parameter values such as the rate of infection or incubation rate shift over the time period of the simulation due to mutations in the virus or public policy that impact behavior.

        \end{enumerate}

    \item[Section~\ref{sec:conclusions}:] Contains the concluding remarks and comments on future directions currently under investigation. 

\end{description}



\section{Background}\label{sec:background}

The type of inverse problem that is formulated and the methods developed for solving such a problem are dictated by the assumptions made about the type of uncertainty that is to be quantified. 
Uncertainties are usually categorized as being either aleatoric (i.e., irreducible) or epistemic (i.e., reducible) in the uncertainty quantification (UQ) community.
Below, we provide a brief literature review on the UQ methods developed to tackle the different inverse problems formulated for these two types of uncertainties to help situate the contributions of this manuscript within the vast UQ literature on these topics.
This review is given primarily at a conceptual-level to avoid introducing any unnecessary notation at this point of the manuscript.

Bayesian methods \cite{Kapteyn2021, CKS14, Gelman2013, CDS10, KO2001, BMP+1994, Fitzpatrick1991} are perhaps the most popular means of inferring probabilistic descriptions of model parameters from QoI data.
In a typical Bayesian framework, one of the initial assumptions is that of an additive noise model on the data that follows a given distribution, usually assumed to be Gaussian.
This assumption is used to describe the uncertainty associated with measurement errors that can theoretically be reduced by collecting more data of the same fidelity or by using improved instrumentation to collect more precise data.
In other words, the assumption is fundamentally that uncertainty is epistemic in nature. 
The solution to the resulting inverse problem within the Bayesian framework is known as a posterior.
The posterior is a conditional density defined by the product of a prior density on parameters and a data-likelihood function.
The data-likelihood function is often formulated in terms of products of the density associated with the noise distribution evaluated at residuals constructed from the differences in simulated and observed QoI data. 
The posterior is interpreted as defining the relative likelihoods that a fixed estimate for the parameters of interest could have produced all of the observed (noisy) data.
The maximum a posteriori (MAP) is often used to estimate parameter values~\cite{Burger_2014}.
Much work has been dedicated to the efficient approximation of the MAP point, e.g.,~\cite{PMS+14} and~\cite{APS+16} utilize local Gaussian approximations of the posterior for this purpose.
Under a typical setup and assumptions in the Bayesian framework, the Bernstein-von Mises theorem~\cite{AsymptoticStats1998} guarantees that the posterior becomes more ``spiked'' around the true parameter value and subsequently that the resulting uncertainty in the parameter estimate is reduced.

The data-consistent (DC) framework is based on a measure-theoretic approach to defining the inverse problem and its solution in terms of pullback and push-forward measures \cite{BET+14, BJW18a, BBE24}. 
Specifically, the objective is to construct a probability measure on the parameter space whose push-forward through the QoI map matches the observed probability measure on the QoI values. 
In other words, the DC solution defines a pullback of the observed probability measure.
In recent years, the density-based approximation of the DC solution, as derived in \cite{BJW18a} via the Disintegration Theorem \cite{Chang_Pollard}, has seen the most development, analysis, and application, e.g., see~\cite{ZM2023, rumbell2023novel, MSB+22, BWZ22, tran2021solving, BGW2020, BH20}.
It is worth noting that a similar form of the density-based DC solution was also derived earlier in \cite{PR2000} through heuristic arguments based on logarithmic pooling and referred to as ``Bayesian melding.'' 
The common thread in these works is that the uncertainty is considered aleatoric due to natural and irreducible uncertainties in both the parameters and the associated QoI data.
This distinction from the typical Bayesian assumption of epistemic uncertainty led to a distinction of the terminology used in the DC framework in \cite{BJW18b} (which is a follow-up to \cite{BJW18a}).
In \cite{BJW18b} and many of the works that chronologically follow it, an initial and predicted density are used to describe the initial quantification of uncertainties on parameters and QoI, respectively, independent of any observed data.
The observed density describes the quantification of uncertainty for the observed QoI data.
An update to the initial density is then obtained via the product of the initial density with the ratio of observed to predicted densities evaluated on the outputs of the QoI map. 
The updated density serves as the DC solution.

While \cite{BYW20} extended the DC framework to handle problems that simultaneously involve both aleatoric and epistemic uncertainties, the contributions of this current work are mathematically most related to \cite{pilosov2023parameter}, which extended the DC framework specifically to solve UQ problems involving epistemic uncertainty in parameters by utilizing the maximal updated density (MUD) point estimate.
In \cite{pilosov2023parameter}, the full linear Gaussian theory is provided along with comparisons to the Bayesian MAP and least squares estimates.
It is shown that the utilization of the predicted density in the construction of the update results in a ``selective regularization'' in the sense that the initial density only impacts the position of the MUD estimate in directions not informed by the data. 
Moreover, it is shown that data-derived QoI map from the residuals of simulated and observed data reduces the variance in MUD parameter estimates as more data are included. 
A quantitative diagnostic originally developed in \cite{BJW18a} for assessing a key predictability assumption in the construction of the updated density is utilized in \cite{pilosov2023parameter} to also assess the quality of the data-derived QoI map.
We build upon all of these aspects in this work to develop the novel algorithm presented in Section~\ref{subsec:seq_alg}. 
We therefore focus the rest of this section on summarizing, briefly, the precise mathematical content of \cite{BJW18a} and \cite{pilosov2023parameter} to set the stage for the contributions of this current work.
For a more thorough comparison of DC and Bayesian frameworks and methods, we direct the interested reader to Sections 2 and 4 of \cite{pilosov2023parameter}, Section 7 in \cite{BJW18a}, Sections 1 and 2 of \cite{BYW20}, and to a recent review paper \cite{BBE24}.

\subsection{Terminology and Notation}\label{sec:dci-background}

Denote by $\pspace$ the space of (input) parameters for the model.
Denote by $Q$ the (potentially vector-valued) QoI map from the parameter space, $\pspace$, to the QoI space defined by $\qspace:= \set{Q(\lambda)\, : \, \lambda\in\pspace}$.
We utilize the notation $\qspace$ instead of $\mathcal{D}$ found in earlier works such as \cite{BJW18a, pilosov2023parameter} to emphasize that this space is distinct from the space of (observational or predicted) data on model outputs since such data are utilized to learn the QoI map.
For simplicity in presentation, we assume $\pspace \subseteq \mathbb{R}^p$ and $\qspace \subset \mathbb{R}^q$ for finite $p$ and $q$.
We use the measure-theoretic shorthand $Q^{-1}(E)$ for any $E\subset\qspace$ to denote the {\em pre-image} of $E$, i.e., $Q^{-1}(E)=\set{\lambda\in\pspace \, : \, Q(\lambda)\in E}$.
Unless otherwise specified, we assume that $\pspace$ and $\qspace$ are equipped with (Borel) $\sigma$-algebras to define measurable spaces, $Q$ is a measurable map between these spaces, and that subsets of these spaces are taken from these $\sigma$-algebras.

Given an observed probability measure, denoted by $\obsmeas$, on $\qspace$, a DC solution is defined by any probability measure $\mathbb{P}_\pspace$ on $\pspace$ such that
\begin{linenomath*}    
\begin{equation}\label{eq:data-consistent}
\mathbb{P}_\pspace(Q^{-1}(E)) = \obsmeas(E), \ \forall E\subseteq\qspace.
\end{equation}
\end{linenomath*}
The DC solution is non-unique unless the map $Q$ is a bijection.
Utilizing an initial probability measure, denoted by $\initmeas$, on $\pspace$, along with a disintegration theorem \cite{Chang_Pollard} and a predictability assumption (defined in Section~\ref{sec:pred_assump} below), \cite{BJW18a} derived the following density-based DC solution that scales well with increasing parameter dimension and is stable with respect to perturbations in the initial and observed probability measures, 
\begin{equation}\label{eq:updens}
    \updens(\lambda) := \initdens(\lambda)\frac{\obsdens(Q(\lambda))}{\preddens(Q(\lambda))}.
\end{equation}
Here, the $\initdens$ and $\obsdens$ are the densities (or, more generally, Radon-Nikodym derivatives) of $\initmeas$ and $\obsmeas$, respectively, while $\preddens$ is referred to as the predicted density that is associated with the push-forward of $\initmeas$ through the QoI map.
The DC solution, $\updens$, is then referred to as the updated density.
We often define the ratio
\begin{equation}\label{eq:ratio}
    r(Q(\lambda)):= \frac{\obsdens(Q(\lambda))}{\preddens(Q(\lambda))},
\end{equation}
which represents the mismatch of observed to predicted relative likelihoods in a QoI value associated with a particular parameter, and rewrite~\eqref{eq:updens} as
\begin{equation}\label{eq:updens_r}
    \updens(\lambda) := \initdens(\lambda)r(Q(\lambda)). 
\end{equation}

Let $\pmb{q}\in\qspace$ denote a fixed QoI value. 
It is clear from~\eqref{eq:updens_r} that if $\lambda\in Q^{-1}(\pmb{q})$, then the ratio $r(Q(\lambda))=r(\pmb{q})$ is constant.
Subsequently, the conditional likelihoods of $\lambda\in Q^{-1}(\pmb(q))$ are identical for both the updated and initial densities.
This is referred to as ``selective regularization'' in \cite{pilosov2023parameter}.



\subsection{Quantitative Diagnostic and the Predictability Assumption}\label{sec:pred_assump}

Generally, $\initdens$ is specified independently from the observed data to represent {\it a priori} assumed uncertainties in parameter values, and $\obsdens$ is defined from observational QoI data. 
It follows that $\updens$ fundamentally relies upon the ability to construct or estimate $\preddens$. 
Moreover, if $q<p$ (i.e., $Q$ maps from a higher-dimensional parameter space to a lower-dimensional QoI space), then the DC solution is defined by the solution to a lower-dimensional forward UQ problem.
We seek to exploit this in the sequential algorithm of Section~\ref{subsec:seq_alg} by iterating over lower-dimensional QoI maps learned from the data.
It is therefore necessary to quantitatively assess the reliability of using any particular estimate of $\preddens$ for a given QoI map to construct $\updens$.
We utilize a quantitative diagnostic originally designed to evaluate the {\it predictability assumption} defined in \cite{BJW18a}.

In measure-theoretic terms, in order for $\updens$ to exist as a density (or Radon-Nikodym derivative) on $\pspace$, $\obsmeas$ must be absolutely continuous with respect to $\predmeas$.
While this is a theoretically sufficient ``predictability assumption'' to guarantee $\updens$ exists in the form given in~\eqref{eq:updens}, computational approaches such as rejection sampling schemes require a stronger form of this assumption (see~\cite{BJW18a}) that we formally define below.

\begin{definition}[Predictability Assumption]\label{def:pred}
    $\exists$ $C>0$ such that $\obsdens(\pmb{q})\leq C\preddens(\pmb{q}))$ for a.e.~$\pmb{q}\in\qspace$.
\end{definition}

If this predictability assumption is met, then $\updens$ defines a density. 
It immediately follows that:
\begin{equation}\label{eq:diagnostic}
    \mathbb{E}_\text{init}(r(Q(\lambda))) = \int_\pspace r(Q(\lambda))\, d\initmeas = \int_\pspace \initdens(\lambda)r(Q(\lambda)\, d\pmeas = \int_\pspace \updens(\lambda)\, d\pmeas = 1,
\end{equation}
where $\pmeas$ denotes the dominating measure (often taken to be the Lebesgue measure although this is not a requirement) on $\pspace$. 
Thus, if the predictability assumption is met, then given a set of independent identically distributed (iid) parameter samples drawn from the initial distribution, the corresponding
sample average of the ratio $r(Q(\lambda))$ should approximate unity. 
To make this a computationally cheap diagnostic to compute, we generally re-use the same parameter samples involved in estimating $\preddens$.
Specifically, we often estimate $\preddens$ from a set of QoI samples generated by evaluating the QoI map on an iid set of parameters generated from the initial distribution. 
Since the QoI on such a set of samples are already available, the quantitative diagnostic defined by the sample average of $r(Q(\lambda))$ can be evaluated without requiring any further model simulations.
This diagnostic proves to be invaluable in assessing the reliability of the updated density and the validity of any statistical inferences drawn from it when utilizing the algorithms presented in Section~\ref{sec:seq-estimation}.

\subsection{Parameter Estimation with MUD Points and Learned QoI maps}




The maximal updated density (MUD) estimate is defined as
\begin{equation}\label{eq:mudpt_inital_defn}
	\lambda^\text{MUD} := \underset{\lambda}{\arg\max} \ \updens(\lambda),
\end{equation}
where $\updens$ is the updated density given in~\eqref{eq:updens} or~\eqref{eq:updens_r}.
Following the analysis of existence and uniqueness of MUD points for linear QoI maps with Gaussian initial and observed distributions \cite{pilosov2023parameter}, the notion of data-constructed QoI maps  is explored. 
The goal is to learn a QoI map that aggregates residuals of observed and simulated data so that the updated density has the property that as more data are utilized to learn the QoI map, the updated covariance around a MUD point shrinks in directions informed by the data. 
Below, we summarize the approach for learning such a QoI map for the general case of non-linear measurement maps.

Suppose that $\lambda^\dagger$ denotes the true parameter value associated with observed data of true system states denoted by $\pmb{z}^\dagger$.
Here, for simplicity of notation, we assume that $\pmb{z}^\dagger$ denotes all system states across any domain of space and time for which the system is observed.
Due to practical limitations in observability, we assume the observed data are determined by taking a finite number of (spatio-temporal) measurements of the states that are polluted by an additive noise model. 
We denote this data by $\set{d_{j}}_{j=1}^n$, which is mathematically defined as
\begin{equation}\label{eq:noisy_data}
    d_{j} := \mathcal{M}_j(\lambda^\dagger; \pmb{z}^\dagger) + \xi_j, \ \ \xi_j\sim \mathcal{N}(0,\sigma_j^2), \ \ 1\leq j\leq n.
\end{equation}
Here, $\xi_j$ denotes the noise in the $j$th datum and $\mathcal{M}_{j}$ denotes the mathematical operator associated with the $j$th measurement device.
Assume we propagate an iid set of $k$ samples, denoted by $\{\lambda^{(i)}\}_{i=1}^k$, drawn from the initial distribution $\initdens$, through the model to generate an associated ensemble of simulated states $\set{\pmb{z}^{(i)}}_{i=1}^k$.
We then form a residual matrix $X \in \mathbb{R}^{k \times n}$ between the observations and the predicted values $\mathcal{M}_j(\lambda^{(i)}, \pmb{z}^{(i)})$, which is defined element-wise as
\begin{linenomath*}
\begin{equation}\label{eq:matrix_res}
	X_{ij} = \frac{\mathcal{M}_j(\lambda^{(i)}, \pmb{z}^{(i)})- d_{j}}{\sigma_{j}}.
\end{equation}
\end{linenomath*}
Performing a principal component analysis (PCA) \cite{pearson1901liii, hotelling1933analysis, jolliffe2002principal, shlens2014tutorial} on this residual matrix and retaining the top $q$ components that explain the most variance in the $n$-dimensional data cloud defined by the rows of $X$, we learn the QoI map, $Q_{PCA}: \pspace \rightarrow \mathbb{R}^q$, where the $\ell$th component of this map evaluated on the $i$th parameter sample is given by
\begin{linenomath*}
\begin{equation}\label{eq:q_pca} 
	(Q_\text{PCA})_{\ell}(\lambda^{(i)}) := \sum_{j=1}^n \pmb{p}^{(\ell)}_j \frac{\mathcal{M}_j(\lambda^{(i)}, \pmb{z}^{(i)}) - d_j}{\sigma_j}, \quad 1 \leq \ell \leq q.
\end{equation}
\end{linenomath*}
Here, $\pmb{p}^{(\ell)}_j$ denotes the $j$th component of the $\ell$th principal component (vector), denoted $\pmb{p}^{(\ell)}$, of $X$. 
It is worth noting that the $Q_\text{PCA}$ map computes a weighted average of residuals, with the weights determined by the principal components of $X$. 
This is utilized in the analysis of \cite{pilosov2023parameter} to demonstrate that this QoI map possesses favorable properties for solving the parameter estimation problem, including:

\begin{enumerate}
    \item Each component of $Q_\text{PCA}$ is a normalized (in the 2-norm) combination of the Z-scored residuals. It follows that the marginal of $\obsdens$ for each $Q_\text{PCA}$ component follows an $\mathcal{N}(0,1)$ distribution. 
    Subsequently, if $\mathcal{M}_j$ is linear for each $j$, then for any fixed value of $q$, the variance in the MUD estimate is reduced in $q$-directions in $\pspace$ as $n$ (the number of data) increases. 
    \item The ordering of the residual matrix $X$ is irrelevant since the PCA results do not depend on column order (i.e., the same QoI map is produced by re-ordering the sum in~\eqref{eq:q_pca}). Thus, it does not matter in what order the data are collected or indexed.
\end{enumerate}

Choosing the appropriate $q$ in the PCA analysis is the critical issue.
If the $n$ data are sensitive to all $p$ of the parameters and $n>p$, then we generally seek to use $q=p$ principal components in constructing the QoI map. 
We explain this at both a mathematical and a conceptual level by considering the simplified case where $\mathcal{M}_j$ is linear for each $j$ and the noise is suppressed in each datum.
With these simplifying assumptions, the rank of $X$ cannot be more than $p$ by standard linear algebra results.
Letting $q\leq p$ denote the rank of $X$ in this case, the $Q_\text{PCA}$ map can be written as $A\lambda+\pmb{b}$ where $A\in\mathbb{R}^{q\times p}$ has orthogonal rows and $\pmb{b}\in\mathbb{R}^q$ is a bias term constructed from sums of the scaled data as evident by~\eqref{eq:q_pca}.
In this case, if $\lambda^\text{init}$ and $\Sigma_\text{init}$ denote the mean and covariance, respectively, of an initial distribution given by $\mathcal{N}(\lambda^\text{init},\Sigma_\text{init})$, then the linear Gaussian theory provided in \cite{pilosov2023parameter} applies so that $\updens\sim \mathcal{N}(\lambda^\text{MUD}, \Sigma_\text{update})$, where
\begin{equation}\label{eq:mud-point-linear-analytical}
    \lambda^\text{MUD} = \lambda^\text{init} + \Sigma_\text{init} A^\top \Sigma_\text{pred}^{-1}(-\pmb{b} -A\lambda^\text{init}),
\end{equation}
and the covariance associated with this point is given by\begin{equation}\label{eq:updatedCov_final}
    \Sigma_\text{update} = \Sigma_\text{init} - \Sigma_\text{init}A^\top \Sigma_\text{pred}^{-1}\left[\Sigma_\text{pred}-\Sigma_\text{obs}\right]\Sigma_\text{pred}^{-1}A\Sigma_\text{init},
\end{equation}
where $\Sigma_\text{pred} := A\Sigma_\text{init}A^T$.
As shown in~\cite{pilosov2023parameter}, $\lambda^\text{MUD}$ exists on the $(p-q)$-dimensional hyperplane in $\pspace$ defined by the intersection of the $q$ orthogonal $(p-1)$-dimensional hyperplanes associated with the nullspaces of $(Q_\text{PCA})_\ell$ for $1\leq\ell\leq q$, i.e., the nullspaces defined by the $q$ rows of $A$.
The position of $\lambda^\text{MUD}$ on the $(p-q)$-dimensional hyperplane is determined entirely by the initial distribution as seen by substituting the definition of $\Sigma_\text{pred}$ into~\eqref{eq:mud-point-linear-analytical}.
If $q=p$, then $A$ becomes a square orthogonal matrix so that $A^\top A$ is the $p\times p$ identity matrix, and the initial distribution no longer plays a role in determining $\lambda^\text{MUD}$.
In the more typical case with nonlinear measurements and noisy data, we seek the largest $q$ possible (up to $p$) and utilize the diagnostic $\mathbb{E}_\text{init}(r(Q_\text{PCA}(\lambda)))$ as an important measure of the quality of the updated density obtained with such a QoI map, and thus the reliability of $\lambda^{MUD}$.
We can systematically reduce $q$ if the diagnostic ever indicates that unreliable QoI were learned from the map, which can occur in the presence of a significant magnitude of noise.
That is, low signal-to-noise ratios may result in learning a $Q_\text{PCA}$ map with certain components explaining variance primarily due to the noise in the observed data, instead of explaining how the variation in simulated data differs from the observed data due to the variation in parameter values. 
This is particularly relevant when developing methods for sequential parameter estimation since data in certain time windows may not exhibit significant sensitivity to perturbations of particular parameters. 
This is fundamental to the main contributions of this paper as we present in detail in Section~\ref{subsec:seq_alg} and is critical in the wind drag example of Section~\ref{sec:adcirc}.

\subsection{Change Point Identification (CPI)}\label{subsec:cpi}
We conclude this section with a brief discussion on Change Point Identification (CPI) Problems.
At a high-level, a CPI problem is one where we need to detect the shift in the true parameter value over time.
Thus, this discussion also serves to highlight the interplay between the two primary contributions of this manuscript highlighted in Section~\ref{sec:intro}. 

CPI is a vital component of various data analysis problems, including economics, infectious disease modeling, and industrial control. 
It requires detecting the point in a sequence of time series data where there is a change in the underlying model or parameters driving the system. 
Although there is a considerable amount of statistical literature on single and multiple change-point models (e.g., see \cite{Adams2007-un},\cite{Stephens1994-bd},\cite{Barry1993-iu}), this current work focuses solely on the CPI problem for the multiple point cases. 

Let $t_0$ denote the initial time at which model simulations begin and $\lambda^{\dagger,0}$ denote the true parameter vector at the beginning of the model simulations.
As before, we denote by $\set{\pmb{d}^{(i)}}_{i=1}^n$ the (noisy) data vectors (possibly of different dimensions) collected at times $\set{t_i}_{i=1}^n$ where $t_0<t_1<t_2<\cdots<t_n$.
Denote by $\set{\tau_j}_{1}^m$ the $m$ times $t_0 < \tau_1< \tau_2 <\cdots < \tau_m < t_n$ where at least one component of the true parameter vector changes values, and denote by $\lambda^{\dagger, j}$ the true parameter vector starting at time $\tau_j$ for $1\leq j\leq m$, respectively. 
We assume that there is not more than one $\tau_j$ between any $t_i<t_{i+1}$ (i.e., not more than one parameter shift occurs between consecutive observation times).
As before, we assume that data are delivered in packets so that all data collected over a time window become available for analysis.
If $\lambda^{\dagger, j}$ is constant over consecutive time windows, then the goal is to produce a sequence of improved MUD estimates for this true parameter value as the data packets become available.
By improved estimates, we mean that both the pointwise accuracy in the estimate should improve as well as $\updens$ becoming more concentrated over the true parameter value with each update. 
If $\lambda^{\dagger, j}$ shifts to $\lambda^{\dagger, j+1}$ over a given time window, then the goal is to detect this shift via the diagnostics and adjust any assumptions about the initial distribution on a given time window as a result so that the MUD estimates shift from estimating $\lambda^{\dagger, j}$ to $\lambda^{\dagger, j+1}$.

\section{Sequential Data-Consistent Parameter Estimation}\label{sec:seq-estimation}

\subsection{The Foundational Algorithms}\label{sec:dci_algs}

Here, we provide an overview of practical issues involving the computation of the DC update, MUD estimate, and the diagnostics critical to evaluating and controlling aspects of the sequential algorithm presented in Section~\ref{subsec:seq_alg}.
Two algorithms are presented in this current subsection that serve as the building blocks for the sequential algorithm.

For a given $\initdens$, QoI map $Q$, and $\obsdens$, estimating $\updens$ at a given $\lambda\in\pspace$ reduces to  estimating the ratio $r(Q(\lambda))$ defined in~\eqref{eq:ratio}.
It follows that evaluating $r(Q(\lambda))$ requires some estimation of $\preddens$. 
Even in situations where $\initdens$ and $\obsdens$ are specified or estimated as belonging to some parametric family of distributions (e.g., Gaussian distributions), it is still often the case that $\preddens$ and thus $\updens$ are both non-parametric. 
In this work, for the sake of both simplicity and reproducibility of results, we estimate $\preddens$ with standard kernel density estimation (KDE) \cite{Devroye1985} techniques.
For a standard KDE estimate of $\preddens$, the basic idea is to take a set of iid QoI samples $\set{\pmb{q}^{(i)}}_{i=1}^k\sim\preddens$ and compute the estimate $\widehat{\preddens}$ defined as
\begin{equation}\label{eq:KDE_pred}
    \widehat{\preddens}(\pmb{q}) := \frac{1}{k} \sum_{i=1}^k \prod_{j=1}^q K_{h_j}\left(\pmb{q}_j, \pmb{q}^{(i)}_j\right)
\end{equation}
where $K_{h_j}$ and $h_j$ define, respectively, the kernel function and bandwidth parameter used for the $j$th dimension in $\qspace$.
Before we address the choice of kernel function and bandwidth parameter, we discuss the practical issue of constructing the iid QoI samples that follow the predicted distribution.
To obtain such a set of samples, we typically generate $\set{\lambda^{(i)}}_{i=1}^k\sim\initdens$ and evaluate the QoI map so that $\pmb{q}^{(i)} = Q(\lambda^{(i)})$ for $1\leq i\leq k$, which is an iid sample from $\preddens$ by construction.
However, due to the sequential estimation we consider in this work, we are often confronted with the situation where a set of QoI samples are associated with parameter samples that are \emph{not} drawn from the initial distribution specified at a given iteration. 
In such cases, we use a \emph{weighted} KDE (wKDE) to construct $\widehat{\preddens}$ defined as
\begin{equation}\label{eq:wKDE_pred}
    \widehat{\preddens}(\pmb{q}) := \frac{1}{\sum_{i=1}^k w^{(i)}} \sum_{i=1}^k w^{(i)}\prod_{j=1}^q K_{h_j}\left(\pmb{q}_j, \pmb{q}^{(i)}_j\right),
\end{equation}
where $w^{(i)}$ denotes the (nonnegative) weight associated with the $i$th sample. 
Note that~\eqref{eq:wKDE_pred} reduces to~\eqref{eq:KDE_pred} if $w^{(i)}=1$ for all $1\leq i\leq k$, which implies that we can always utilize~\eqref{eq:wKDE_pred} as long as the weights are appropriately identified.
In Section~\ref{subsec:seq_alg}, we discuss the selection of these weights in the context of the sequential algorithm. 
As far as the choice of kernel function and bandwidth parameters utilized in this work, we choose the Gaussian kernel function and Scott's rule for choosing the bandwidth parameter that are the default options encoded within the \verb|gaussian_kde| function in the Python library \verb|scipy| \cite{2020SciPy-NMeth}.
These are perhaps the most popular choices in the literature and generally considered robust although we note that other types of kernels and bandwidth selection criteria are also popular, e.g., see \cite{Heidenreich2013}.

Approximating $\preddens$ with $\widehat{\preddens}$ allows for the trivial computation of $r(Q(\lambda))$ and subsequently $\updens$ at these same parameter samples, $\set{\lambda^{(i)}}_{i=1}^k$. 
It is worth noting that some recent works by distinct groups of researchers consider alternative methods of estimating $\updens$.
Such methods include the use of Generative Adversarial Networks (GANs) \cite{rumbell2023novel} and Sequential Monte Carlo (SMC) \cite{rumbell2023sequential}. 
These alternative approaches appear quite promising in addressing approximation issues that plague KDEs in high-dimensions although they still typically require a large number of observed and predicted samples used in the training steps.
A full quantitative comparison of these methods and how they impact the resulting MUD estimates is beyond the scope of the current work and is therefore left for future research. 
However, we emphasize that a key feature of the DC solution is that estimating $\preddens$ immediately produces an estimate of $\updens$.
Thus, in the context of sequential iteration over \emph{low-dimensional} QoI maps, we effectively mitigate concerns regarding high-dimensional spaces when applying KDE techniques as we later demonstrate. 



Given a set of iid parameters $\set{\lambda^{(i)}}_{i=1}^k\sim \initdens$, the corresponding set $\set{Q(\lambda^{(i)})}_{i=1}^k\sim\preddens$, and the subsequent KDE approximation of $\preddens$, we use the sample average of $\mathbb{E}_\text{init}(r(Q(\lambda))$, defined in~\eqref{eq:diagnostic}, to determine (i) if any violation of the predictability assumption occurs, (ii) if the QoI map learned from data is suitable for constructing the DC update, or (iii) if the KDE approximation of $\preddens$ is sufficiently accurate.
In this work, we augment this powerful diagnostic with another quantitative metric based on the Kullback-Leibler (KL) divergence \cite{SNW:kl, SNW:renyikl}, which is a quantification of the expected information gain from changing one distribution to another.
Specifically, \cite{BJW18a} observed that
\begin{linenomath*}
\begin{align}
    \mathrm{KL}_\mathrm{DCI} := \mathrm{KL}\left(\updens \, : \,  \initdens\right) &:= \int_{\Lambda} \updens(\lambda) \log \left(\frac{\updens(\lambda)}{\initdens(\lambda)}\right) \, d\pmeas \\
    &= \int_{\Lambda} r(Q(\lambda)) \log (r(Q(\lambda))) \, d\initmeas \label{eq:KL_r}\\
    &= \int_{\qspace} \obsdens(q) \log \left(\frac{\obsdens(q)}{\preddens(q)}\right) \, d\qmeas \\
    &=: \mathrm{KL}\left(\obsdens \, : \, \preddens \right) \label{eq:KL_DCI}.   
\end{align}
\end{linenomath*}
it follows that the information gained by solving the DC inverse problem (i.e., replacing $\initdens$ with $\updens$) is exactly the information gained by replacing $\preddens$ with $\obsdens$.
In other words, solving a forward UQ problem is sufficient for determining the information gained by solving the DC inverse problem, which is the basis for an efficient DC-based optimal experimental design framework studied in \cite{WWJ17}. 
In this work, we find this equivalence useful in the context of sequential parameter updates for problems that may exhibit parameter drift, which we discuss in more detail later in this section. 
For now, we remark that the form given in~\eqref{eq:KL_r} is simply a sample average (with respect to the initial distribution) of $r(Q(\lambda))\log(r(Q(\lambda)))$.
It follows that it is straightforward to estimate this as an output diagnostic similar to how we estimate $\mathbb{E}_\text{init}(r(Q(\lambda))$.


\begin{algorithm}
    \caption{Weighted Data-Consistent Inversion (wDCI)
    }
    \label{alg:dci_weighted}
    \begin{algorithmic}[1]
    \Function{\textbf{wDCI}}{$S_\text{DC}$ as defined in Eq.~\ref{eq:s_dc_state}}
        \State $\preddens \leftarrow \textbf{wKDE}\left(\set{\pmb{q}^{(i)}}_{i=1}^k, \set{w^{(i)}}_{i=1}^k \right) $ \Comment{Weighted KDE on forward model evaluations, cf.~Eq.~\eqref{eq:wKDE_pred}.}
        \State $\set{r^{(i)}}_{i=1}^k \leftarrow \set{\frac{\obsdens(\pmb{q}^{(i)})}{\preddens(\pmb{q}^{(i)})}}_{i=1}^k$ \Comment{Compute update ratios for each sample, cf.~Eq.~\eqref{eq:ratio}}
        \State $\mathbb{E}_\text{init}(r) \leftarrow \frac{1}{k}\sum_{i=1}^{k}r^{(i)}w^{(i)}$ \Comment{Sample average diagnostic (should be $\approx 1$).}
        \State $\mathrm{KL}_\mathrm{DCI} \leftarrow \sum_{i=1}^k w^{(i)}r^{(i)}\log(r^{(i)})$ \Comment{Compute information gain diagnostic, cf.~Eq~\eqref{eq:KL_r}.}
        \State \Return $\left\{\preddens, \set{r^{(i)}}_{i=1}^k, \mathbb{E}_\text{init}(r), \mathrm{KL}_\mathrm{DCI}\right\}$
    \EndFunction
  \end{algorithmic}
\end{algorithm}

Algorithm~\ref{alg:dci_weighted} summarizes the (weighted) DC inversion (wDCI) computations and diagnostics considered in this work.
The inputs to this algorithm define the necessary components of the state of the modeled system to perform wDCI

\begin{equation}\label{eq:s_dc_state}
    S_\text{DC}=\left\{
    \begin{aligned}
        &\set{\pmb{q}^{(i)}}_{i=1}^k \subset \qspace &&: \text{Forward model evaluations } Q(\lambda^{(i)}) \text{ on a given sample of parameters} \set{\lambda^{(i)}}_{i=1}^k. \\
        &\set{w^{(i)}}_{i=1}^k \subset\mathbb{R}^+ &&: \text{Optional weights for each parameter sample. Defaults to $w_i = 1 \ \forall 1\leq i \leq k$}. \\
        &\obsdens &&: \text{Observed distribution on $\qspace$.}
    \end{aligned}
    \right\}
\end{equation}

As mentioned above, we often generate $\pmb{q}^{(i)}$ by evaluating the QoI map on a set of iid samples $\lambda^{(i)}\sim\initdens$ for $1\leq i\leq k$, in which case $w^{(i)}=1$ for all $1\leq i\leq k$.
If the parameter samples are not drawn according to the distribution defined by $\initdens$, then the weights are defined by $w^{(i)}=\initdens(\lambda^{(i)})$ for all $1\leq i\leq k$. 
The outputs of this algorithm are $\preddens$, the ratios $\set{r^{(i)}}_{i=1}^k:=\set{r(\pmb{q}^{(i)})}_{i=1}^k$, and the sample average estimates of the two diagnostics $\mathbb{E}_\text{init}(r)$ and $\mathrm{KL}_\mathrm{DCI}$.

\algrenewcommand\algorithmicfunction{}
\begin{algorithm}
	\caption{Maximal Update Density (MUD) Parameter Estimation
    } \label{alg:mud-pca}
    \begin{algorithmic}[1]
        \Function{\textbf{MUD}}{$S_\text{MUD}$ as defined in Eq.~\ref{eq:s_mud_state}}       
            \State $X_{ij} \leftarrow \sigma_j^{-1}(\mathcal{M}_{ij}- d_j)$ \Comment{Compute z-scored residual matrix, cf.~\eqref{eq:matrix_res}.}
            \State $\set{\pmb{p}^{(\ell)}}_{\ell=1}^q \leftarrow PCA(X)$   \Comment{Compute first $q$ principal components of residual matrix.}
            \State $\set{\pmb{q}^{(i)}}_{i=1}^k \leftarrow \set{(Q_{PCA})(\lambda^{(i)})}_{i=1}^k$ \Comment{Use $\set{\pmb{p}^{(\ell)}}_{\ell=1}^q$ and ~\eqref{eq:q_pca} to construct QoI samples.} 
            \State $S_\text{DC}\leftarrow \set{\set{\pmb{q}^{(i)}}_{i=1}^k, \set{w^{(i)}}_{i=1}^k, \mathcal{N}(\pmb{0},I_{q\times q})}$ \Comment{Define wDCI state.}
            \State $\left\{\preddens, \set{r^{(i)}}_{i=1}^k, \mathbb{E}_\text{init}(r), \mathrm{KL}_\mathrm{DCI}\right\} \leftarrow \text{\bf wDCI}\left(S_\text{DC}\right)$ \Comment{Use Algorithm~\ref{alg:dci_weighted}.}
            \State $\lambda^\text{MUD} \leftarrow \lambda^{(i_{\max})}$ for $i_{\max} := \text{argmax}_{i} w^{(i)} r^{(i)}$ \Comment{Determine MUD estimate.}
            \State \Return $\left\{\preddens, \set{r^{(i)}}_{i=1}^k, \mathbb{E}_\text{init}(r), \mathrm{KL}_\mathrm{DCI},  \lambda^\text{MUD}\right\}$ 
        \EndFunction
    \end{algorithmic}
\end{algorithm}

Algorithm~\ref{alg:mud-pca} summarizes the MUD parameter estimation computations including the role of the wDCI algorithm (i.e., Algorithm~\ref{alg:dci_weighted}) in this estimation.
Much like Algorithm~\ref{alg:dci_weighted}, the inputs to this algorithm define the necessary components for the state of uncertainty for the modeled system to perform MUD estimation:

\begin{equation}\label{eq:s_mud_state}
    S_\text{MUD} = 
        \left\{
            \begin{aligned}
                &\set{d_j}_{j=1}^n, \set{\sigma_j}_{j=1}^n  &&: \text{Observed (noisy) spatio-temporal data and variance in noise, Eq.~\eqref{eq:noisy_data}} \\
                &\set{\lambda^{(i)}}_{i=1}^k \subset\pspace &&: \text{Parameter samples (not necessarily drawn from $\initdens$).} \\
                &\set{w^{(i)}}_{i=1}^k \subset\mathbb{R}^+ &&: \text{Optional weights for each parameter sample. Defaults to $w_i = 1 \ \forall 1\leq i \leq k$}. \\
                &\mathcal{M}_j(\lambda^{(i)}, \pmb{z}^{(i)}) =: \mathcal{M}_{ij} &&: \text{Predicted/simulated measurements associated with parameters.} \\
                & q &&:  \text{Number of QoI to construct from principal components.}  
            \end{aligned}
        \right\}
\end{equation}

While all the computations in Algorithm~\ref{alg:dci_weighted} take place in the space $\qspace$ to perform generic wDCI for updating an (unspecified) initial density, Algorithm~\ref{alg:mud-pca} involves computations in both $\pspace$ and $\qspace$.
Subsequently, $S_\text{MUD}$ explicitly involves the parameter samples whereas $S_\text{DC}$ does not.
The remaining inputs defining $S_\text{MUD}$ are necessary for constructing the $Q_\text{PCA}$ map and components needed for defining $S_\text{DC}$ as described above. 

\subsection{The Sequential MUD Estimation Algorithm}\label{subsec:seq_alg}

At a high-level, sequential MUD estimation is straightforward to explain with a few small adjustments to existing notation.
First, denote by $\set{\pmb{d}^{(i)}}_{i=1}^N$ the (noisy) data vectors (possibly of different dimensions due to availability of certain measurements at different times\footnote{Consider, for example, data obtained by an orbiting satellite for a particular spatial domain on the earth, which is only periodically available and may be reduced in spatial fidelity due to cloud cover.}) collected at times $\set{t_i}_{i=1}^N$ where $t_0<t_1<t_2<\cdots<t_N$.
Mathematically, we assume the $j$th component the $i$th data vector, denoted by $\pmb{d}^{(i)}_j$, is of the form shown in~\eqref{eq:noisy_data}. 
The MUD estimate associated with analyzing all data simultaneously is obtained as follows: (i) concatenate all the data vectors into a single vector of dimension $n=\sum_{i=1}^N \text{dim}(\pmb{d}^{(i)})$, (ii) define $\set{d_i}_{i=1}^n$ as the set of data defined by the $n$ components of this concatenated vector, and (iii) execute Algorithm~\ref{alg:mud-pca} as usual.
This approach has several drawbacks including having to wait until the final data are collected to produce a parameter estimate.
We therefore present a sequential approach to MUD estimation that assumes packets of data are made available semi-regularly throughout time in Algorithm~\ref{alg:seq_MUD}.
After the collection of a data-packet, a candidate MUD solution is constructed, assessed for quality by checking necessary diagnostics, and either accepted and propagated forward (by re-sampling or re-weighting of parameter samples) or rejected.
If the diagnostics suggest to reject a solution, new candidate MUD solutions may be constructed and re-tested against the diagnostics.

At a more detailed level, let $\set{t_{i_m}}_{m=1}^M$ denote the $M\leq N$ data-transmission times where $t_{i_0}:=t_0<t_1\leq t_{i_1} < \cdots < t_{i_{M-1}} < t_n \leq t_{i_M}$.
We assume that the data collected are stored locally within the measurement system network, all local data are transmitted at the specified data-transmission time, and then the locally stored data are purged from the network memory. 
In other words, at time $t_{i_m}$, the data packet is defined by all the data collected {\em after} (but not including) $t_{i_{m-1}}$ and up to (and including) $t_{i_m}$. 
At a given data-transmission time $t_{i_m}$, the goal is to update the current state of uncertainty and any existing MUD estimate based on the data collected in the time window $(t_{i_{m-1}},t_{i_m}]$ for $1\leq m\leq M$.
To this end, let $\initdens^{(m)}$ denote the initial density assumed for time window $m$ defined by $(t_{i_{m-1}},t_{i_m}]$ for $1\leq m\leq M$.
We use all data collected in this time window to construct a {\em candidate} for the $m$th MUD estimate and updated density via Algorithm~\ref{alg:mud-pca}, denoted by $\lambda^{\text{MUD}, m}$ and $\updens^{(m)}$, respectively.
Upon constructing a candidate solution, Algorithm~\ref{alg:seq_MUD} checks the corresponding diagnostics to determine how to proceed.
Options include (1) retrying (lines 9-14) with either a lower-dimensional learned QoI map (Control 1), increasing the number of parameter samples (Control 2), or deploy a new set of samples or sample weights (Controls 3 and 4), (2) using the solution in the next iteration (lines 16-21) via either re-sampling or re-weighting, or (3) skipping the iteration all together.

In summary, the inputs to Algorithm~\ref{alg:seq_MUD} can then be defined as:

\begin{equation}\label{eq:s_smud_state}
    S = 
            \left\{
                \begin{aligned}
                    &\initdens^{(1)}  &&: \text{Initial distribution over parameters.} \\
                    &\set{\pmb{d}^{(i)}}_{i=1}^N, \set{t_i}_{i=1}^N  &&: \text{Observed data, and observation times} \\
                    &\set{\lambda^{(i)}}_{i=1}^k \subset\pspace &&: \text{Parameter samples (not necessarily drawn from $\initdens^{(1)}$)} \\
                    &\set{w^{(i),1}}_{i=1}^k \subset\mathbb{R}^+ &&: \text{Optional initial parameter weights. Defaults to $w_i = 1 \ \forall 1\leq i \leq k$}. \\
                    &\set{t_{i_m}}_{m=1}^M &&: \text{Data transmission times} \\
                    &\epsilon_\text{pred}, \epsilon_\text{KL} &&: \text{Tolerances for diagnostics} \\
                    & q &&:  \text{Number of QoI to construct from principal components.}
                \end{aligned}
            \right\}
\end{equation}
Below, we provide mathematical context for utilizing the diagnostics in Algorithm~\ref{alg:seq_MUD} to control certain aspects Algorithms~\ref{alg:dci_weighted} and~\ref{alg:mud-pca} (which we refer to below simply as the wDCI and MUD algorithms, respectively) on each iteration.

\subsubsection{Diagnostic 1: Solution Validity}\label{subsec:diag1}

Recall the first diagnostic verifies the validity of the predictability assumption by comparing an estimation of $\er$ to unity.
Provided the estimate of $\er$ is within a user-specified threshold of unity denoted by $\epsilon_{pred}$, we consider the candidate MUD solution as valid and simply proceed with carrying the updated information into the MUD estimation for the next data packet.
Assuming the predictability assumption holds for an exact $\preddens$, we note that choosing an appropriate value for $\epsilon_{pred}$ for evaluating the suitability of an approximation to $\preddens$ is problem dependent with the learned QoI map dimension and the sample size being the greatest factors impacting the accuracy in the density estimate. 
Therefore, in this work, we control for three factors (QoI dimension, sample size, and parameter drift) that can cause the diagnostic to indicate a potential violation of the predictability assumption.

Control 1 (line 10) reduces the dimension of the constructed QoI map, which can reduce approximation errors in the estimation of $\preddens$ with a fixed finite sample size.
Moreover, this helps to limit the analysis to the components of the QoI map that exhibit the greatest sensitivity to parameters.
Alternatively, we may increase the sample size by opting for Control 2 (line 11) to directly reduce the approximation errors in the KDE of the given QoI map although this requires additional forward model evaluations.
If Controls 1 and 2 fail, we essentially rule out approximation error of $\preddens$ due to QoI dimension or sample size as the issue. 
If we can also rule out parameter drift (discussed below in Section~\ref{subsec:diag3}), then we conclude that the data are not sensitive to the parameters at a given iteration.
This points to a critical issue in parameter estimation problems for dynamical systems since it is important to distinguish between ``bad'' data and a ``bad'' model.



\algrenewcommand\algorithmicfunction{}
\begin{algorithm}
  \caption{Sequential MUD (sMUD) Parameter Estimation
  }
  \label{alg:seq_MUD}
  \begin{algorithmic}[1]
    \Function{\textbf{sMUD}}{$S$ as defined in Eq.~\ref{eq:s_smud_state}}
        \For{$m = 1; m\leq M; m \mathrel{+}= 1$}
            \State $\pmb{d}\leftarrow \text{concatenate } \pmb{d}^{(i)} \text{ for } i_{j-1}<i\leq i_j$ \Comment{Aggregate data in $(t_{i_{j-1}},t_{i_j}]$ into single data vector}
            \State $\set{\sigma_j}_{j=1}^n \leftarrow$ Measurement variances for $\set{\pmb{d}_j}_{j=1}^n$,  $n=\text{dim}(\pmb{d})$ \Comment{Get measurement device statistics}
            \State $\set{M_{ij}}_{1\leq i\leq k, 1\leq j\leq n} \leftarrow \set{\mathcal{M}_j(\lambda^{(i)}, \pmb{z}^{(i)})}_{1\leq i\leq k, 1\leq j\leq n}$ \Comment{Simulate $n$ data for $k$ samples}
            \State $S_{\text{MUD},m} \leftarrow  \set{\set{\pmb{d}_j}_{j=1}^n, \set{\sigma_j}_{j=1}^n, \set{\lambda^{(i)}}_{i=1}^k,  \set{w^{(i),m}}_{i=1}^k, \set{M_{ij}}_{1\leq i\leq k, 1\leq j\leq n}, q}$ \Comment{$m$th MUD state}
            \State $\left\{\preddens^{(m)}, \set{r^{(i)}}_{i=1}^k, \mathbb{E}_\text{init}(r), \mathrm{KL}_\mathrm{DCI}, \lambda^{\text{MUD}, m} \right\} \leftarrow \textbf{MUD}\left(S_{\text{MUD},m} \right)$  \Comment{Obtain MUD estimate}
            \If {$\abs{\mathbb{E}_\text{init}(r)-1}\geq \epsilon_\text{pred}$} { choose from control options:}
                \If {$\mathrm{KL}_\mathrm{DCI} < \epsilon_\text{KL}$}
                    \State {\bf Control 1:} Return to line 6 and set $q\leftarrow q-1$ in MUD state
                    \State {\bf Control 2:} Return to line 5 adding $\set{\lambda^{(i)}}_{i=k+1}^{k+\Tilde{k}}$ new samples from $\initdens^{(m)}$.
                \Else { potential shift detected}
                    \State {\bf Control 3:} Return to line 6 using new $\initdens^{(m)}$ to re-weight samples  $\set{w^{(i),m}}_{i=1}^k=\set{\initdens^{(m)}(\lambda^{(i)})}_{i=1}^k$
                    \State {\bf Control 4:} Return to line 5 using new $\initdens^{(m)}$ with re-sampled parameters $\set{\lambda^{(i)}}_{i=1}^k\sim\initdens^{(m)}$
                \EndIf
            \Else { accept candidate solution}
                \If {$k_{\text{eff}}/k < \epsilon_{samples}$} {weight collapse}
                    \State $\initdens^{(m+1)} \leftarrow \textbf{wKDE}\left(\set{\lambda^{(i)}}_{i=1}^k, \set{w^{(i),m}r^{(i)}}_{i=1}^k \right), \set{w^{(i),m+1}}_{i=1}^k \leftarrow 1 $ \Comment{Update initial explicitly}
                    \State $\set{w^{(i),m+1}}_{i=1}^k \leftarrow 1 $ \Comment{Re-set weights}
                    \State Draw new samples $\set{\lambda^{(i)}}_{i=1}^k \sim \initdens^{(m+1)}$ \Comment{Draw new samples for next iteration}
                \Else
                    \State $\set{w^{(i),m+1}}_{i=1}^k \leftarrow \set{w^{(i),m}r^{(i)}}_{i=1}^k$ \Comment{Update initial implicitly via re-weighting of samples}
                \EndIf
            \EndIf
        \EndFor
        \State \Return $\set{\set{\lambda^{\text{MUD},m}}_{m=1}^M, \set{\set{w^{(i),m}}_{i=1}^k}_{m=1}^M}$ \Comment{Return sequence of MUD estimates}
    \EndFunction
  \end{algorithmic}
\end{algorithm}

\subsubsection{Diagnostic 2: Weight Collapse}\label{subsec:diag2}

Having accepted a candidate MUD solution, we must choose one of two options for proceeding to the next iteration: re-sampling or re-weighting.
Ideally, we simply re-weight the existing parameter samples with the new weights on the following iteration.
This allows us to avoid generating new samples and running the associated simulations for these new samples as this can involve complex and tedious ``hot-starting'' for complex computational simulation codes.
However, it is possible for the weights to ``collapse'' over multiple iterations in the sense that the weights become concentrated around a small set of parameter values. 

To mathematically see this, consider how for a given data packet obtained at time $m$, we construct the learned $q$-dimensional QoI map, denoted by $Q_\text{PCA}^{(m)}$ within the MUD algorithm, along with the predicted density, denoted by $\preddens^{(m)}$, from the embedded call to the wDCA algorithm. 
This implies that each of the $M$ updated densities, denoted by $\set{\updens^{(m)}}_{m=1}^M$, is given by
\begin{equation}
    \updens^{(m)}(\lambda) = \initdens^{(m)}(\lambda)\frac{\obsdens^{(m)}(Q_\text{PCA}^{(m)}(\lambda))}{\preddens^{(m)}(Q_\text{PCA}^{(m)}(\lambda))} = \updens^{(m-1)} \frac{\obsdens^{(m)}(Q_\text{PCA}^{(m)}(\lambda))}{\preddens^{(m)}(Q_\text{PCA}^{(m)}(\lambda))} = \initdens^{(1)}\prod_{k=1}^m \frac{\obsdens^{(k)}(Q_\text{PCA}^{(k)}(\lambda))}{\preddens^{(k)}(Q_\text{PCA}^{(k)}(\lambda))}, \ 1\leq m\leq M.
\end{equation}
In other words, the $m$th updated density is not only an update of the $m$th initial density but is also an update of the first initial density. 
This hints at the ``weight collapse" issue that can impact the $\mathbb{E}_\text{init}(r(Q(\lambda))$ diagnostic that is computed at each iteration.
Suppose the parameter samples are generated from $\initdens^{(1)}$ and are held fixed throughout all the simulations.
If the updated densities begin to drift substantially away from $\initdens^{(1)}$ (as measured by the KL divergence), then it is possible that only a few parameters will have most or all of the weight as defined by the $w^{(i)}$ values computed in the MUD algorithm at a particular iteration.
In this case, the wKDE estimates of the predicted densities will become inaccurate and $\mathbb{E}_\text{init}(r(Q(\lambda))$ will deviate significantly from unity in later time windows. 

To reduce these errors due to weight collapse, we define the effective sample size on any iteration as
\begin{equation}
    k_{\text{eff}}^{(i)} := \#\left\{i \in \{1, 2, \ldots, k\} \mid \frac{w^{(i),1}}{\sum_{j=1}^k w^{(j),1}} > \epsilon_{\text{mach}}\right\},
\end{equation}
i.e., the number of (normalized) weights that are greater than some machine epsilon (usually around $1e-16$, i.e. non-zero), and then only propagate the new sample weights forward if $k_{\text{eff}}^{(i)}/k > \epss $ where $0 < \epss \leq 1$.
Note that by setting $\epss = 1$, the algorithm is forced to choose re-sampling while setting $\epss < 0.5$ runs the risk of propagating a poorly resolved updated distribution into the next iteration, especially for smaller sample sizes. 

\subsubsection{Diagnostic 3: Change-Point Identification (CPI)}\label{subsec:diag3}

The final diagnostic considered in the sMUD Algorithm~\ref{alg:seq_MUD} relates to identifying parameter drift as the potential cause for a predictability assumption violation. 
The DC information gain given by $\kl$ in \eqref{eq:KL_DCI} can correlate breakdowns in the predictability assumption with large shifts in the updated distribution that indicate potential parameter drift outside of the range of the current sample set.
Possible actions in response to this are to (1) reset the parameter sample weights or (2) reset the initial distribution and drawing new samples.

To understand how the sMUD algorithm detects a potential drift, recall from Section~\ref{subsec:cpi} that in a CPI problem we want to detect when $\lambda^{\dagger, j}$ shifts to $\lambda^{\dagger, j+1}$ in a time window $(t_{i_{m-1}},t_{i_m}]$.
Consider the simple fact that if $\lambda^{\dagger,j}$ is constant over many time windows, then the updated densities have a tendency to concentrate around the $\lambda^{\text{MUD}, m}$ estimates, which are themselves converging (such behavior is shown in the numerical examples).
If there is a drift in the true parameter, the subsequent predicted densities (which involve push-forwards of the concentrated updated densities) over the next few time windows are likely to be concentrated away from the observed density resulting in a detected violation of the predictability assumption.
Note how during the execution of the sMUD algorithm, we are able to distinguish this particular root cause of a breakdown in the predictability assumption from the approximation errors discussed in Section~\ref{subsec:diag1} because (1) Controls 1 and 2 will fail to improve estimates, and (2) the DCI information gain $\kl$ will be greater than in preceding iterations, indicating a large shift in the updated distribution.

We therefore define a threshold $\epskl$ so that $\kl > \epskl$ is flagged as a potential parameter drift and we choose either from Control 3 or 4 to modify the initial distribution and samples to re-solve the problem in the given time window by searching in regions where the shifted true parameter $\lambda^{\dagger, j+1}$ now belongs.
The particular choice of control to take during a parameter drift is problem dependent.
The key factor for choosing the appropriate control is to ensure that the support of the new initial distribution of parameter samples covers a set containing the new true parameter.
Note that it is possible for a parameter drift to occur and not trigger one of the controls if the updated distribution has not yet become sufficiently concentrated around the prior true parameter value $\lambda^{\dagger, j}$ so that their exists samples deemed likely by their associated weights in the range of the current true parameter value $\lambda^{\dagger, j+1}$. 

The CPI applications of the sMUD algorithm are studied in more depth in the final numerical example (see Section~\ref{sec:seir}) where the inclusion of the KL divergence also provides correlation of poor $\mathbb{E}(r)$ values with how much information gain is present in the update.
Specifically, we demonstrate that when the expected ratio breaks the $\epsilon_\mathrm{pred}$ threshold due to sampling and/or measurement error, the KL divergence between the update and the predicted is small.
However, when a parameter shift occurs, the deviation of the expected ratio is due to a shift in dynamics, and therefore also a significant parameter update and large KL divergence are observed.

\section{Numerical Results}\label{sec:results}

The general framework of the sMUD Algorithm~\ref{alg:seq_MUD} allows for its application to a variety of different scenarios that differ in (1) data ingestion strategies (i.e. offline vs online) (2) re-sampling capabilities and (3) parameter input and output dimensions. 
We see how these differing scenarios influence our strategies for choosing the appropriate diagnostic values and decision control statements.
Here, we apply the sequential MUD (sMUD) Parameter Estimation Algorithm~\ref{alg:seq_MUD} to the following problems involving dynamical systems governed by differential equations:

\begin{itemize}

    \item Estimating a 2-dimensional wind-drag parameter for a storm-surge model requiring HPC resources.
    In this case, the computational burden of running the forward model limits our ability to re-sample. 
    With no re-sampling, we are limited on each iteration to Control option 1 (line 10 in Algorithm~\ref{alg:seq_MUD}) to reduce the number of principle components used if the predictability assumption is flagged as potentially violated.
    
    \item Estimating a (spatially-varying time-independent) continuous function representing the thermal diffusivity of a medium that conducts heat.
    The key here is to use quick re-sampling techniques to quickly explore the higher-dimensional parameter space with a relatively small sample size. 
    Thus, all three controls of the problem are available to us on each iteration of the sequential estimation problem.
    
    \item Estimating a 4-dimensional parameter for an epidemiological model with change points  corresponding to the progression of the disease and changing environment.
    All three control statements become relevant, as we need to distinguish between parameter shifts and bad-signal to noise ratio in data packets along with choosing when to re-sample as opposed to re-weighting any given iterative solution using the sMUD diagnostics.
    
\end{itemize}

All the  codes here are produced using the open source pyDCI python library for data-consistent inversion.
See \ref{sec:software} for more information on how to reproduce the results presented here and access supplemental results.

\subsection{Offline Sequential Estimation: Storm Surge}\label{sec:adcirc}

We first apply the sMUD algorithm in an offline sequential estimation scenario, where by ``offline" we mean that we are working with a set of already observed data and simulated samples.
We begin with the same problem set-up as in \cite{pilosov2023parameter} but divide the dataset  into distinct 12-hour data transmission time windows to demonstrate (1) the application of the sMUD algorithm and (2) that similar, if not better, parameter estimates are obtained compared to the original non-sequential approach.
The dataset used in this example is available at \cite{del-castillo-negrete2022shinnecock}.

\subsubsection{ADCIRC with uncertain wind drag}

The system under investigation is governed by the Shallow Water Equations (SWE), which are a widely used depth-averaged approximation of the Navier-Stokes equations.
The SWE are commonly employed in coastal circulation and flooding modeling to accurately predict storm surge resulting from extreme weather events \cite{vreugdenhil1994numerical}.
The mathematical representation of the SWE is as follows:

\begin{align}\label{eq:swe}
\frac{\partial \zeta}{\partial t}+\nabla \cdot(\mathbf{U} H) & =0 \\
\frac{\partial \mathbf{U}}{\partial t}+\mathbf{U} \cdot \nabla \mathbf{U}+f \mathbf{k} \times \mathbf{U} & =-\nabla\left[\frac{p_s}{\rho_0}+g \zeta\right]+\frac{\tau_s-\tau_b}{\rho_0 H}.
\end{align}

The unknown free surface elevation, denoted as $\zeta=\zeta(x, y, t)$, represents the deviation from mean sea level, while the depth-averaged velocity, denoted as $\mathbf{U}=\mathbf{U}(x, y, t)$, is the velocity averaged over the height of the water column. 
Furthermore, $H$ denotes the height of the water column, $f$ represents the Coriolis parameter, $p_s$ represents the atmospheric pressure, $\rho_0$ is the reference density of water, $g$ is the gravitational constant, $\boldsymbol{\tau}_s$ represents the surface stress, and $\boldsymbol{\tau}_b$ represents the bottom stress. 

While there are many drivers of uncertainty in this model, we focus on the uncertainty in the surface stress $\boldsymbol{\tau}_s$. This is commonly modeled as
\begin{align}
\boldsymbol{\tau}_s=\rho_s C_d \mathbf{u}\|\mathbf{u}\|.
\end{align}
The model includes the wind drag coefficient $C_d$, which is an effective parameter that governs the transfer of momentum from winds to the water column. 
This transfer of momentum is one of the primary drivers of storm surge. 
The specific form of $C_d$ depends on the physical properties of the system being modeled, such as the type of storm and presence of ice. 
In this study, a popular generalization of Garratt's formula for $C_d$, as proposed in \cite{garratt1977review} and implemented within ADCIRC, is used where
\begin{align}
C_d=\min \left[10^{-3}\left(.75+\lambda_1 u\right), \lambda_2\right] .
\end{align}
The linear drag law slope parameter $\lambda_1$ is typically set to $0.067$, with a maximum cut-off value $\lambda_2$ typically set to 0.0025 in order to represent the sheeting of waves at high wind speeds ($>27$m/s \cite{holthuijsen2012wind,bryant2016exploration}).
We consider these values to be uncertain in this example.

We solve the SWE with ADCIRC, which is a system of computer programs for solving time dependent, free surface circulation and transport problems in two and three dimensions. 
ADCIRC utilizes a finite-element model of the SWEs, where the Generalized Wave Continuity Equations is discretized in space using piecewise-linear elements on unstructured (triangular) grids \cite{Luettich:1992}.
This model is widely used in coastal engineering applications, such as hurricane storm surge forecasting \cite{dietrich2013realtime}, hindcasting \cite{bunya2010highresolution, dietrich2010highresolution, dietrich2011hurricane}, and uncertainty quantification \cite{BGE+15, graham2015adaptive, GBW+17}.
It can run in both single core and distributed computing environments \cite{tanaka2011scalability,dietrich2012performance}.

We consider the well-tested Shinnecock Inlet test grid, using the same simulated storm set-up as used in \cite{MSB+22, pilosov2023parameter}.
The ADCIRC simulation runs for 16 days, from December 29, 2017 - January 14, 2018, and is forced by tides reconstructed from TPXO9.1 harmonic tidal constituents \cite{egbert2002efficient} using OceanMesh2D \cite{roberts2019oceanmesh2d}, constant air pressure  of 1013 millibars, and hourly 10-m wind velocities at a $0.25^{\circ}$ resolution from the CFSv2 data set \cite{saha2014ncep}.
For the purpose of the studies presented in this work, the winds are scaled artificially by a factor of up to three near the inlet, with the winds smoothly reduced to zero near the outer boundary.

We use a similar setup as in \cite{pilosov2023parameter} and make the initial assumption that the uncertain parameters $\left(\lambda_1, \lambda_2\right)$ fall within a range of $\pm 50 \% $ of commonly used default values of $(0.067, 0.0025)$ mentioned above. 
This defines the finite-dimensional parameter space:
\begin{align}
\Lambda=[0.0335,0.1105] \times[0.00125,0.00375] \subset \mathbb{R}^2 .
\end{align}
We generate 1000 samples from a uniform distribution over $\Lambda$ as inputs to ADCIRC.
Water elevation measurements are recorded at an artificial station inside the inlet over a period of 14 days (1 January 2018-14 January 2018) at intervals of three hours for each sample. 
Since no real station data are available, we create synthetic observations by selecting and removing the sample closest to the default parameter values of $(0.067, 0.0025)$, and adding iid noise from a $\mathcal{N}(0,\sigma^2)$ distribution with $\sigma^2 = 0.1$ to each measurement (as opposed to the original study by \cite{pilosov2023parameter}, which used $\sigma^2 = 0.05$). 

\begin{figure}
    \centering
    \includegraphics[width=\linewidth]{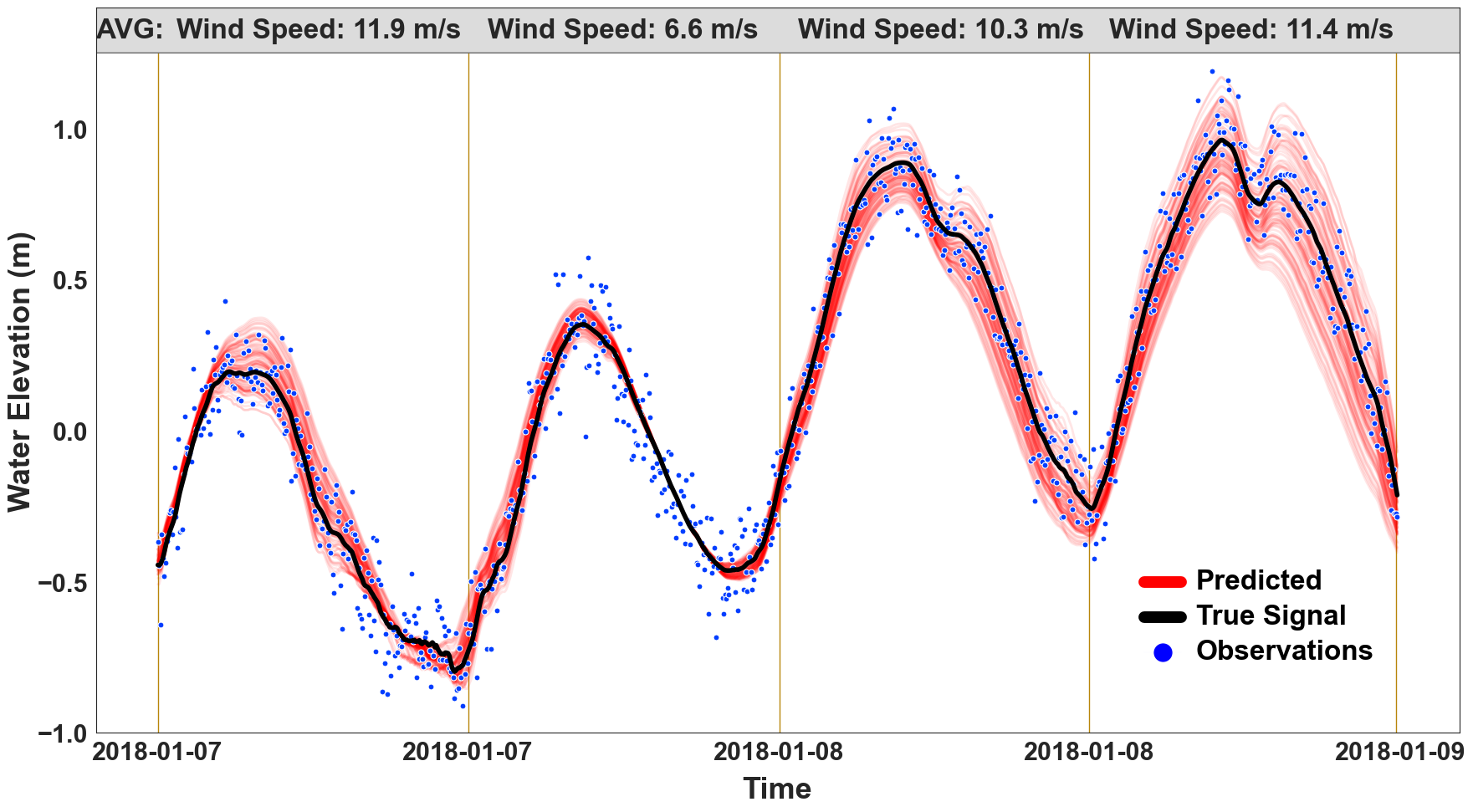}
    \caption{Time Evolution of Water Elevation (Left Axis): True state (black line), Observed state (blue dots), and a sample of 100 Predicted States (red). Vertical lines mark the 12-hour intervals used for transmission times, while the parameter with the highest sensitivity, average wind speed, is located at the top banner. Notably, three time windows exhibit high wind speeds, with the second window featuring lower speeds. This aligns with the most favorable update in the $\lambda_1$ (wind-drag slope) direction due to reduced cut-off parameter dynamics.}
    
    \label{fig:adcirc_ts}
\end{figure}

\subsubsection{12-Hour Data Transimssion Windows}

In \cite{pilosov2023parameter}, it is observed that the choice of time window proves essential to the accuracy and quality of the MUD parameter estimate. 
Specifically, in time windows where the system exhibited low but increasing winds, the slope parameter of the wind drag coefficient, $\lambda_1$, was efficiently estimated while a time window of primarily high winds led to an effective update in the $\lambda_2$ wind-drag cut-off parameter.
A time window that exhibited both high and low winds led to the best MUD estimate with a vector-valued QoI map. 
In this current study, we iterate over a scalar $\qpca$ map to demonstrate how each of these parameters may be systematically estimated over a longer time window as data packets are utilized that exhibit different qualities as the storm progresses. 
Moreover, limiting to a scalar map allows us to more easily explore the evolving geometric relationship between the QoI and the parameter space.

We start with the same time window of data that showed sensitivity to both parameters used in \cite{pilosov2023parameter}, divide this into sequential data transmission time windows, and apply the sMUD algorithm.
In most weather prediction problems (including forecasting storm surges), observational data such as wind speed, central pressure, radius of maximum winds, and water surface elevation are transmitted at intervals ranging from 6 to 12 hours, which motivates the choice of 12-hour time windows in this example, which is depicted in Figure~\ref{fig:adcirc_ts}.

We set the first diagnostic threshold to $\epspr = 0.2$ and note that Controls 1-2 are never taken as each 12-hour time window produces an acceptable candidate solution (see table~\ref{tab:adc}).
We also note that no parameter drift is occurring, so there is no need for Controls 3-4 for the given problem (which is effectively done by setting $\epskl$ to some large number). 
Finally, since we are working with a fixed dataset that is not amenable to resampling, we set $\epss=0.0$ so that the algorithm is forced to use re-weighting at each iteration.

\begin{table}
\centering
\begin{tabular}{|c|c|c|c|}
\hline
\textbf{iteration} & $\er$ & $\kl$ & $k_\text{eff}$ \\
\hline
1 & 1.103141 & 1.469817 & 1.000000 \\
2 & 0.810056 & 1.174189 & 0.563126 \\
3 & 1.066731 & 0.561245 & 0.476954 \\
4 & 0.884539 & 1.585180 & 0.271543 \\
\hline
\end{tabular}
\caption{Diagnostic values over each 12-hour time window for the ADCIRC Parameter Estimation Problem}
\label{tab:adc}
\end{table}

\begin{figure}
    \centering
    \includegraphics[width=\textwidth]{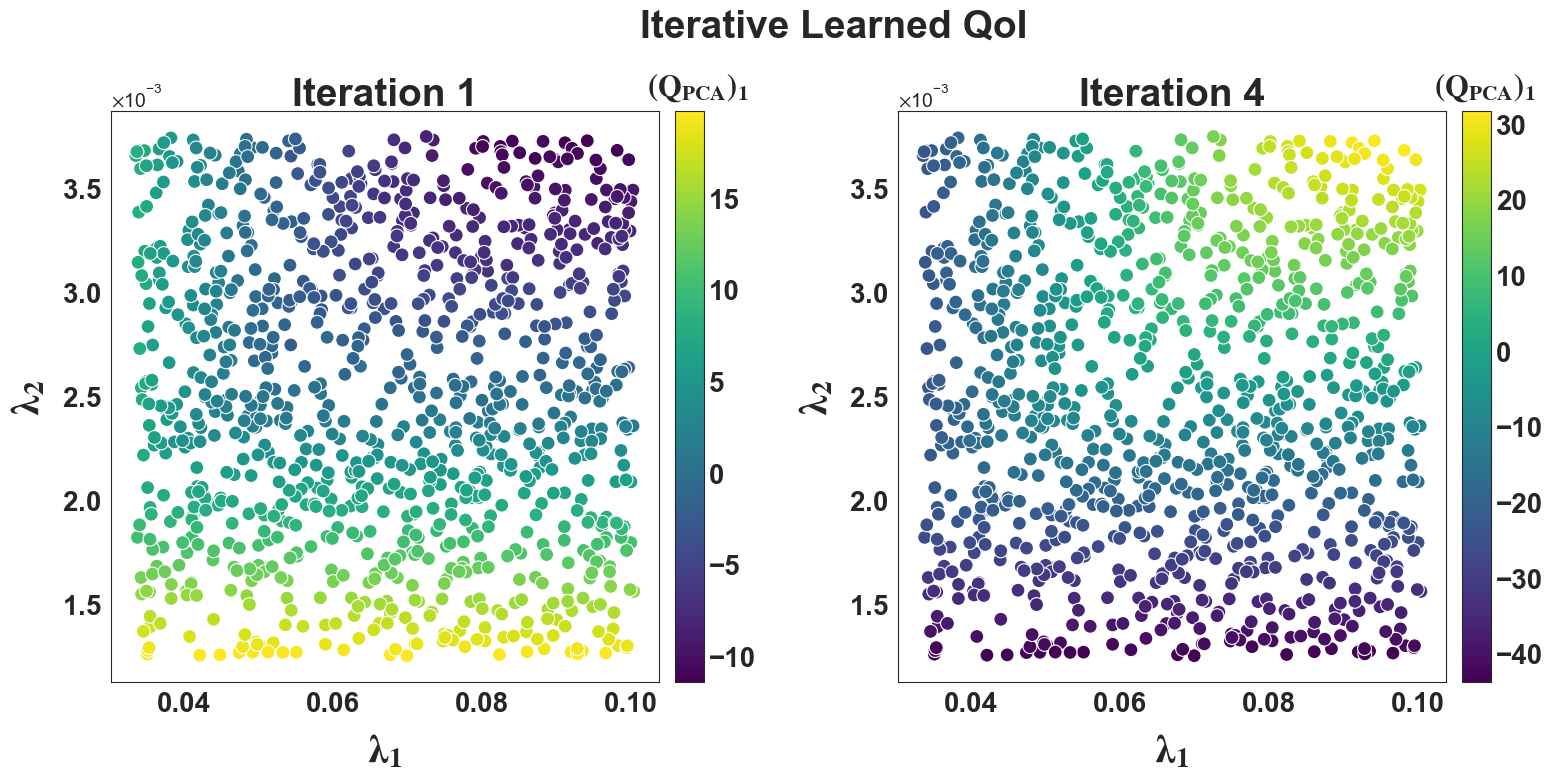}
    \caption{Parameter sample scatter plots, colored by the learned $\qpca$ over the 1st (left) and last (right) iterations of data. Note the difference in implied contour structures across iterations, which illustrates how the sequential estimation allows updating of parameter estimates to occur in distinct directions informed by each time window of data.}
    \label{fig:adcirc_qoi}
\end{figure}

Figure~\ref{fig:adcirc_qoi} shows the distinct geometric structure of the QoI map on two separate iterations of the algorithm (the first and last) as illustrated by a color associated with the QoI value corresponding to each parameter sample. 
By iterating over this scalar map, we solve four computationally cheaper problems in a more operational setting than provided in \cite{pilosov2023parameter}, which analyzed the entire 48-hour time window at once with a vector-valued QoI map. 

\begin{figure}
    \centering
    \begin{subfigure}[b]{\linewidth}
        \centering
        \includegraphics[width=0.8\linewidth]{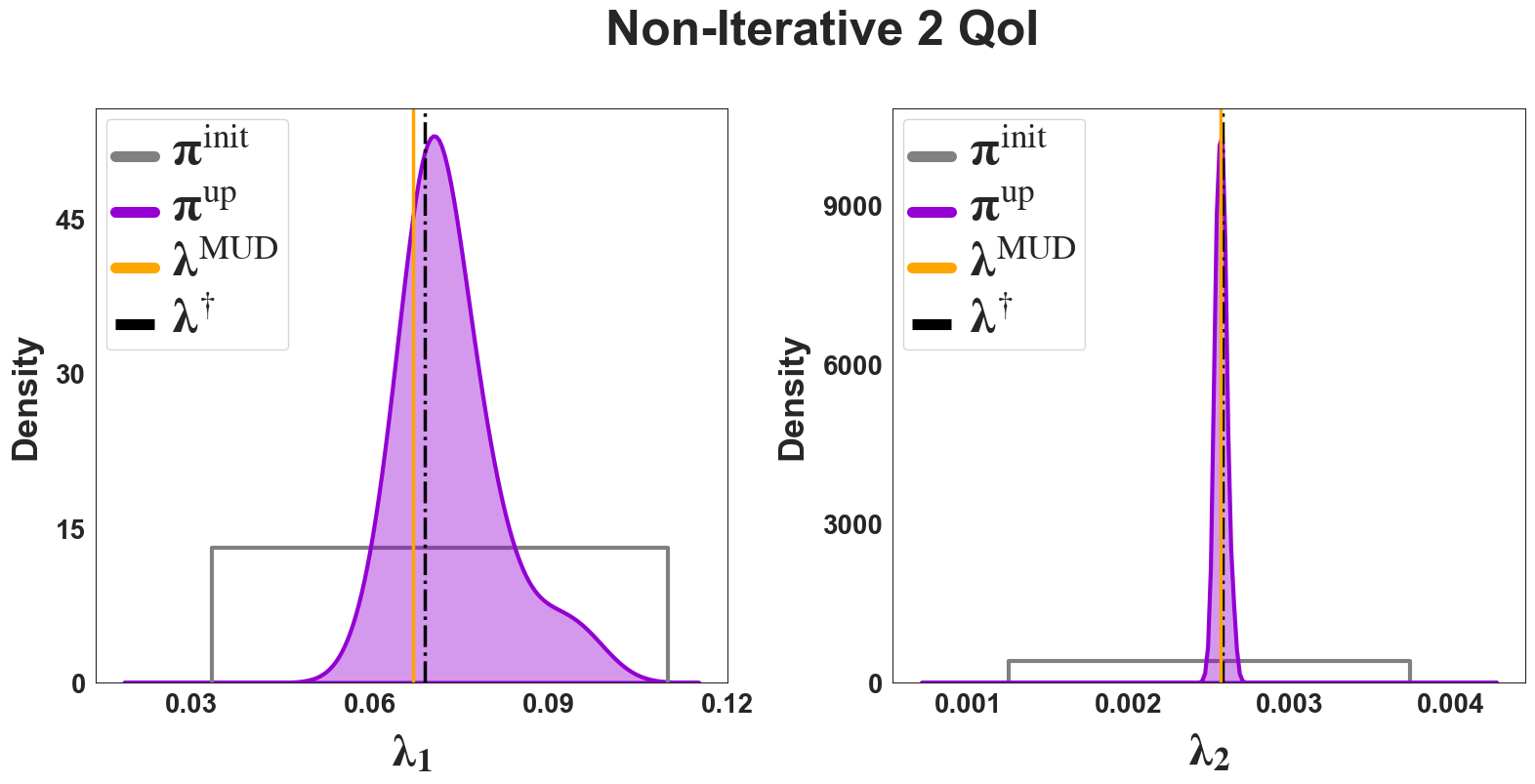}
        \label{fig:adcirc_sol_full}
    \end{subfigure}
    \begin{subfigure}[b]{\linewidth}
        \centering
        \includegraphics[width=0.8\linewidth]{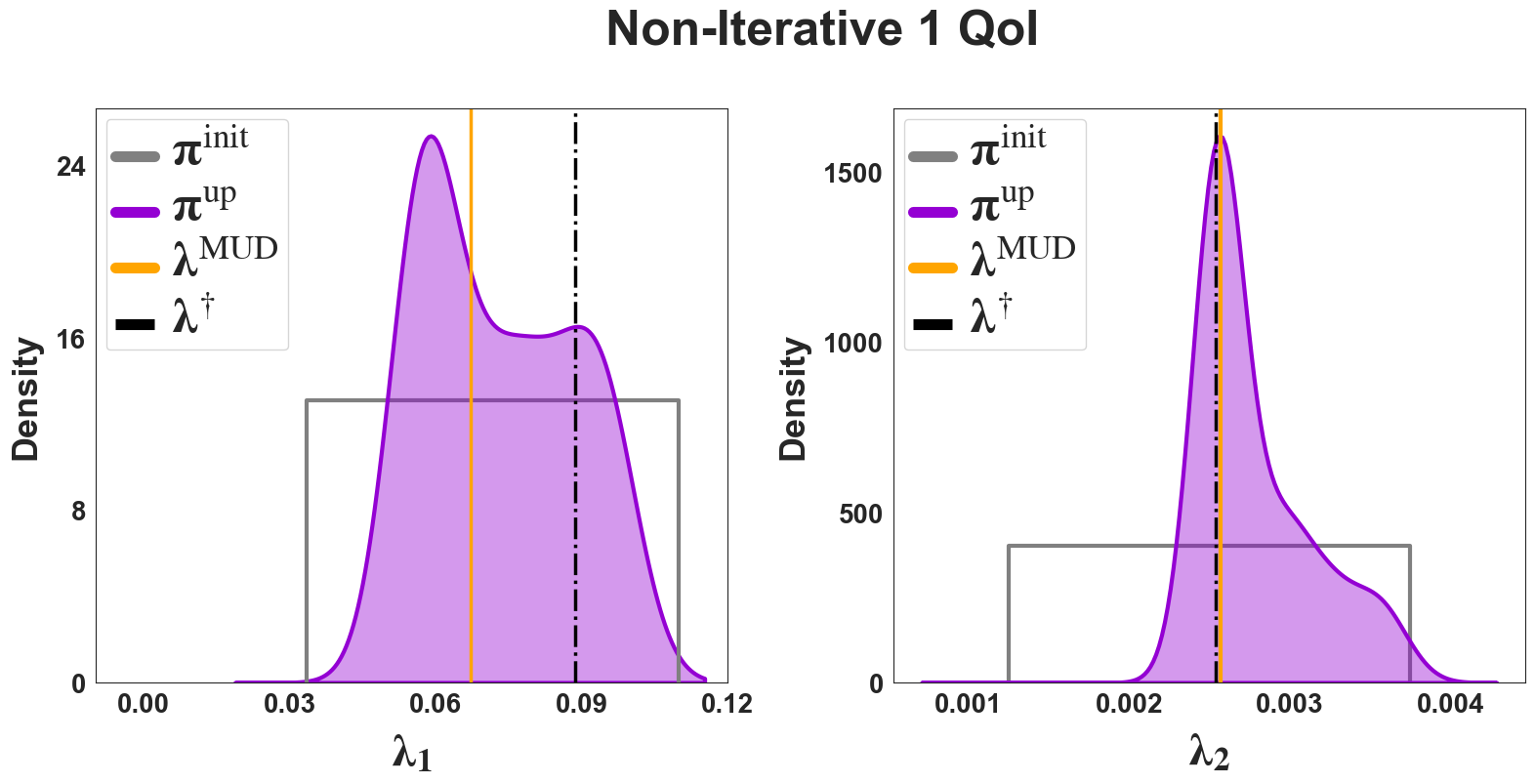}
        \label{fig:adcirc_sol_it2}
    \end{subfigure}
    \begin{subfigure}[b]{\linewidth}
        \centering
        \includegraphics[width=0.8\linewidth]{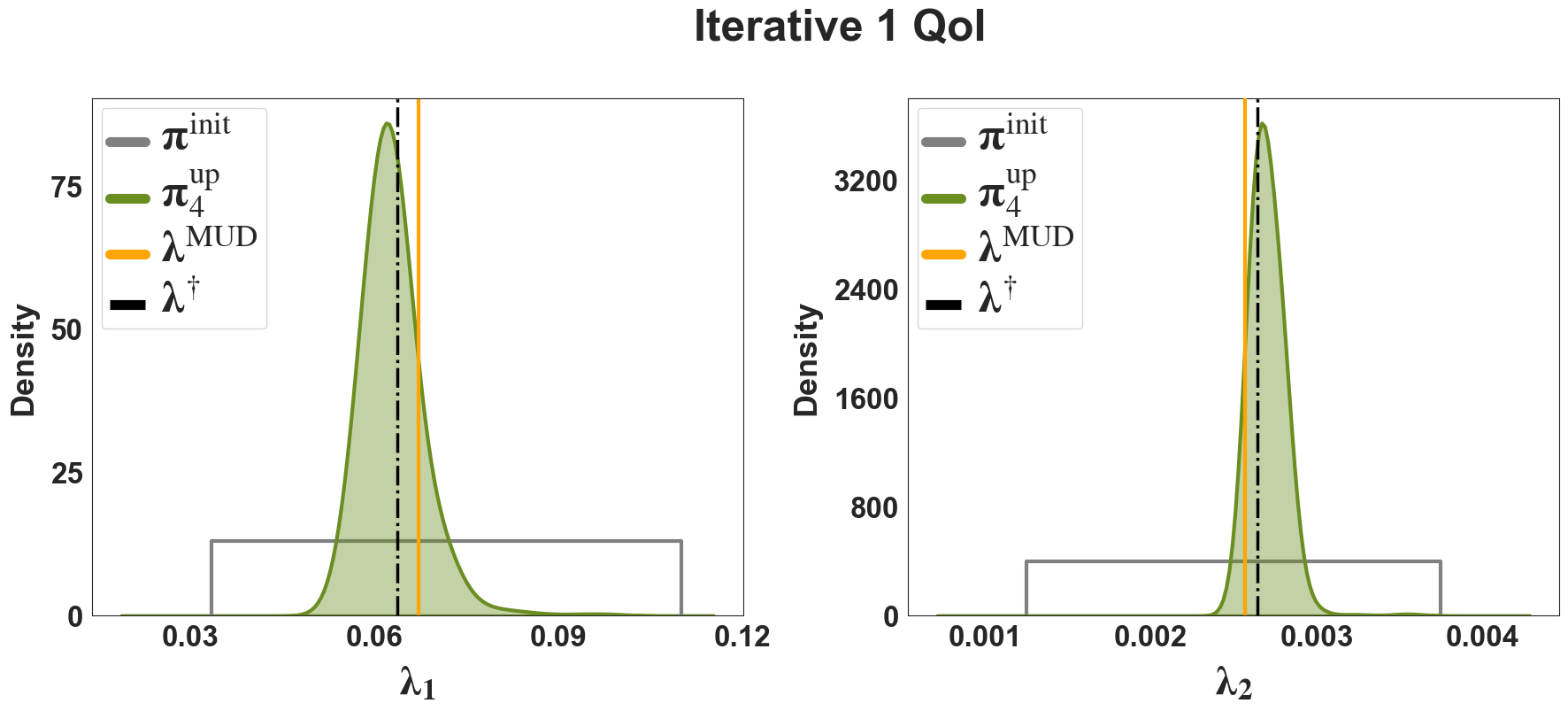}
        \label{fig:adcirc_sol_it1}
    \end{subfigure}
    \caption{ADCIRC Wind Drag Parameter Estimation Results: Initial (black, non-filled in) and updated density plots (purple/green filled in densities) along with true parameter values $\lambda^\dagger$ (vertical black dotted-dash lines) and $\lambda^\mathrm{MUD}$ estimates (vertical orange lines) for the $\lambda_1$ slope parameter (left) and $\lambda_2$ cut-off parameter (right) using three different approaches: (top) Non-iterative, 2 principle component QoI map, (middle) non-iterative 1 principle component QoI map, and (bottom) 12-hour iterative 1 principle component QoI map. Note how the iterative, 1 component map performs similarly well as the non-iterative 2-component map.}
    \label{fig:adcirc_sol1}
\end{figure}

Figure~\ref{fig:adcirc_sol1} further illustrates the comparison of MUD estimates obtained by the sequential approach to that achieved in \cite{pilosov2023parameter}. 
For the non-iterative approach using the full 48-hour time window of data simultaneously, we observe that accurate estimates of the two parameters are only obtained using the two-component $\qpca$ map (Figure~\ref{fig:adcirc_qoi} top) as opposed to the one-component (i.e., scalar) map ((Figure~\ref{fig:adcirc_qoi} middle).
While the MUD estimate in the two-component case appears to be accurate, there is still significant uncertainty in the estimate of $\lambda_1$ compared to $\lambda_2$ as illustrated by comparing the concentrations of the updated marginal densities around these estimates in the left- and right-plots of this top row. 
Comparing these solutions to the iterative solution using 12-hour time windows and a scalar QoI map (Figure~\ref{fig:adcirc_sol1}, bottom), we see that the iterative update produces arguably better results as the uncertainty is more balanced between the estimates of the individual parameters with clearly reduced uncertainty in the MUD estimate for $\lambda_1$ compared to the 48-hour simultaneous analysis of data with a vector-valued map.
We again emphasize that these results are computationally cheaper to obtain as they involve the analysis of (1) less data at each iteration and (2) a reduction in dimension of the QoI map.
Moreover, the sequential approach is more operationally aligned with a real- or near-real time analysis of data and updating of uncertainties since the data in applications such as this are often delivered in 6- or 12-hour intervals. 

\subsection{Online Sequential: The Heat Equation}\label{sec:heat}

We now consider a parabolic partial differential equation (PDE) that is a model  for a wide-range of applications in engineering and physics. 
Here, we interpret this PDE as modeling the diffusion of heat through a given region in a particular medium.
Let $u = u(\mathbf{x}, t) \in  \Omega \times(0, T]$ model the temperature at any point in space $\boldsymbol{x} = (x_{1},x_{2}) \in \Omega$ and time $t \in (0,T]$.
Denote by $k(\mathbf{x}; \lambda) \in \Omega \times \Lambda$ the thermal diffusivity of a given medium, which is considered uncertain in this example and itself modeled as a random field.
Then, denoting the source function as $f \in L^{\infty}(\Omega)$, we have that $u$ satisfies the following PDE, 
\begin{equation}\label{eq:heateq}
\begin{cases}
\begin{aligned}
\frac{\partial u}{\partial t} & =k(\mathbf{x}; \lambda)\nabla^2 u+f & & \text { in } \Omega \times(0, T], \\
u & =u_D  & & \text { in } \partial \Omega \times(0, T], \\
u & =u_0 & & \text { at } t=0 . 
\end{aligned}
\end{cases}
\end{equation}
For simplicity, we prescribe homogeneous Dirichlet boundary conditions $u_{D}=0$ and set $\Omega=[-2,2]\subset \mathbb{R}^d$ for $d=2$. 
To produce datasets with interesting features, we prescribe an initial condition of $u_0 = e^{-5||\mathbf{x}||^2}$ and set the forcing function to be $f = 10\sin(6\pi t) x_{1} + 10\cos(4\pi t) x_{2}$.
With this setup, the goal of the inverse problem is the recovery of the thermal diffusivity $k(\mathbf{x}; \lambda) \in \Omega \times \Lambda$, from noisy measurements of $u$ over time, i.e.
\begin{equation}
\boldsymbol{d} = u(x,t) + \varepsilon, \quad  \varepsilon \sim N\left(\mathbf{0}, \sigma^2 \mathbf{I}\right).
\end{equation}
We use 500 randomly placed sensors over our domain to collect measurements every 0.05 seconds. 

\subsubsection{Parameterizing the Thermal Diffusivity Field via a KL Expansion}

The Karhunen-Loeve (KL) expansion~\cite{huang2001convergence,chada2019tikhonov} is a common tool to parameterize $k(\mathbf{x}; \lambda)$ as:
\begin{equation}\label{eq:kl_expansion}
k(\boldsymbol{x}; \lambda)=\mathbb{E}[k(\boldsymbol{x}; \lambda)]+\sum_{j=1}^{\infty} \sqrt{\lambda_{j}} \xi_{j}(\lambda) \psi_{j}(\boldsymbol{x}),
\end{equation}
where
\begin{equation}\label{eq:kl_coefficients}
\xi_{j}(\lambda)  =\frac{1}{\sqrt{\lambda_{j}}} \int_{\Omega} \left( k(\boldsymbol{x}; \lambda)-\mathbb{E}[k(\boldsymbol{x})]\right)\psi_{j}(\boldsymbol{x}) \mathrm{d} \boldsymbol{x}.
\end{equation}

Note that~\eqref{eq:kl_expansion} allows us to efficiently represent the thermal diffusivity using only the coefficients of the expansion as calculated in~\eqref{eq:kl_coefficients} so that $k(\mathbf{x},\lambda)$ can be considered a Gaussian process consisting of Gaussian iid random variables with zero mean and unit variance. i.e.  $\xi_j(\lambda) \sim \mathcal{N}\left(0,1\right)$. 
Furthermore, we assume $\mathbb{E}[\xi_j \xi_{s}]=\delta_{j {s}}$, where $\xi_i$'s are a set of coordinates that fully characterize the parameter field $k(\mathbf{x},\lambda)$ when the eigenbasis, $\set{\psi_i}$, is known.

In order to calculate the form of the eigenbasis, $\set{\psi_i}$, often referred to as the KL modes, we make the further assumption that  $k(\boldsymbol{x}; \lambda)$ is a square-integrable random field with mean
\begin{equation}
\mathbb{E}[k(\boldsymbol{x}; \lambda)]=\int_{\Lambda} k(\boldsymbol{x}, \lambda) \mathrm{d} \mu_{\boldsymbol{\Lambda}}(\lambda), \quad \boldsymbol{x} \in \Omega
\end{equation}
and a continuous, symmetric, positive definite covariance
\begin{equation}
C\left(\boldsymbol{x}, \boldsymbol{x}^{\prime}\right)=\int_{\Lambda}(k(\boldsymbol{x}, \lambda)-\mathbb{E}[k(\boldsymbol{x})])\left(k\left(\boldsymbol{x}^{\prime}, \lambda \right)-\mathbb{E}[k\left(\boldsymbol{x}^{\prime})\right]\right) \mathrm{d} \mu_{\boldsymbol{\Lambda}}(\lambda) = \sigma(x)\sigma(x^\prime) \exp \left(-\frac{\left(x-x^{\prime}\right)^2}{2 l^2}\right),
\end{equation}
and use Mercer's theorem to calculate the KL Modes
\begin{equation}
\int_{\Omega} \mathcal{C}\left(\boldsymbol{x}, \boldsymbol{x}^{\prime}\right) \psi_{j}
\left(\boldsymbol{x}^{\prime}\right) \mathrm{d} \boldsymbol{x}^{\prime}=\lambda_{j} \psi_{j}
(\boldsymbol{x}), \quad \int_{\Omega} \psi_{j}(\boldsymbol{x}) \psi_{s}(\boldsymbol{x}) \mathrm{d} \boldsymbol{x}=\delta_{j s}.
\end{equation}
Note we assume here some known prior variance structure $\sigma(x)$, which can be spatially varying, and with known length scale.
For our example problem we assume $\mathbb{E}\left[k(x)\right] = 1.0$, $\sigma = 0.2$ is constant over the domain, and a correlation length scale is fixed as $l = 0.1$.

The size of the retained terms $M$ in the KL expansion in~\eqref{eq:kl_expansion} is determined by the desired energy percentage to be retained by the KL expansion, which is defined as $\sum_{j=1}^{M} \lambda_j / \sum_{j=1}^{\infty} \lambda_j$.
In practice, it is recommended to choose the value of $M$ such that the truncated KL expansion captures as much information as possible compared to its infinite counterpart. 
This means that for prior covariance functions with smaller correlation lengths, larger values of $M$ are needed for the KL expansion to capture a similar percentage of information. 
In this example, we take $M = 10$\footnote{The choice of 10 KL modes was made to pose a sufficiently challenging large dimensional problem, yet only requiring modest computational resources to run the forward simulations necessary to solve the problem using the sMUD algorithm. Accompanying code repository (see Appendix~\ref{sec:software}) includes examples of using more/less KL modes, requiring more/less samples and iterations to solve.} and randomly draw 10 values from an $\mathcal{N}(0, 1)$ distribution to determine the set of true parameters (see Figure~\ref{fig:hm_est}).
The goal then reduces to estimating the true thermal diffusivity field as characterized by these ten KL mode coefficients, given {\it a prior} assumption of the covariance as previously described.

One final implementation detail is that in practice we use~\eqref{eq:kl_expansion} to parameterize the log of the thermal diffusivity field. 
This is done to prevent the thermal diffusivity from taking on negative values inconsistent with the problem's physics as a result of the KL expansion.



\subsubsection{Online Sequential Estimation: $T = 0.5s$ time windows}

To solve the parameter estimation problem we apply the sMUD algorithm using $k = 100$ samples, starting from an initial distribution of $\initdens = \mathcal{N}(0, 2)$, with a data transmission time step of $\Delta t = t_{i_m} - t_{i_{m-1}} = 0.5$  seconds for a total time of $M = 6$ steps. 
We use the popular FEniCSx Finite Element library~\cite{baratta2023dolfinx, scroggs2022basix, scroggs2022construction, alnaes2014unified} as our forward model to solve~\eqref{eq:heateq} for each sample.
Synthetic data is collected for the set of reference parameters (the 10 KL mode coefficients chosen to represent the true thermal diffusivity field) by recording measurements at 500 randomly placed sensors over the domain every $t = 0.05$ seconds, and populating those measurements with $\sigma = 0.05$ used to define the noise level.

Applying the sMUD Algorithm~\ref{alg:seq_MUD} successfully to this problem requires a distinct set of parameters than the previous ADCIC problem from section~\ref{sec:adcirc}.
First, despite the much larger parameter space (10 vs 2) and smaller sample size (100 vs 1000), we keep $\epspr$ at 0.2, to ensure that when we only accept good candidate solutions.
On each iteration, by using Controls 1 and 2 to allow for QoI maps ranging from three- to one-dimensional while also allowing for the addition of 50 sample increments to the sample size, we ensure that we are able to obtain a solution and MUD estimate that passes the prescribed diagnostic with $\epspr=0.2$.
Once $\updens$ is obtained at a given iteration, we set the second diagnostic threshold $\epss = 0.9$ to heavily favor re-sampling from the updated distribution at the next iteration since we are working with a smaller sample size and a larger parameter space. 
Setting a high-value for $\epss$ effectively forces each iteration to re-sample $k=100$ new samples and thus forces a faster exploration of the parameter space over the iterations.
Finally, we note that we do not use Controls 3 and 4 in this scenario (by setting $\epskl$ very high) since there is no parameter drift expected.

\begin{figure}
  \centering
  \vspace{-2cm}
   \begin{subfigure}{0.35\textwidth}
    \includegraphics[width=\linewidth]{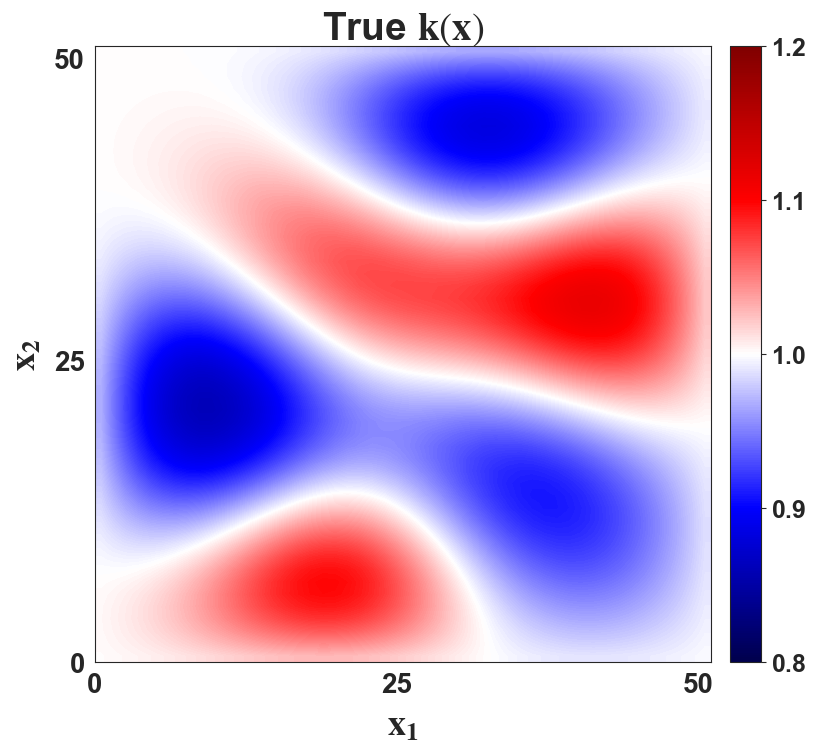}
  \end{subfigure}
  \hspace{0.3\textwidth}
  \begin{subfigure}{0.7\textwidth}
    \includegraphics[width=\linewidth]{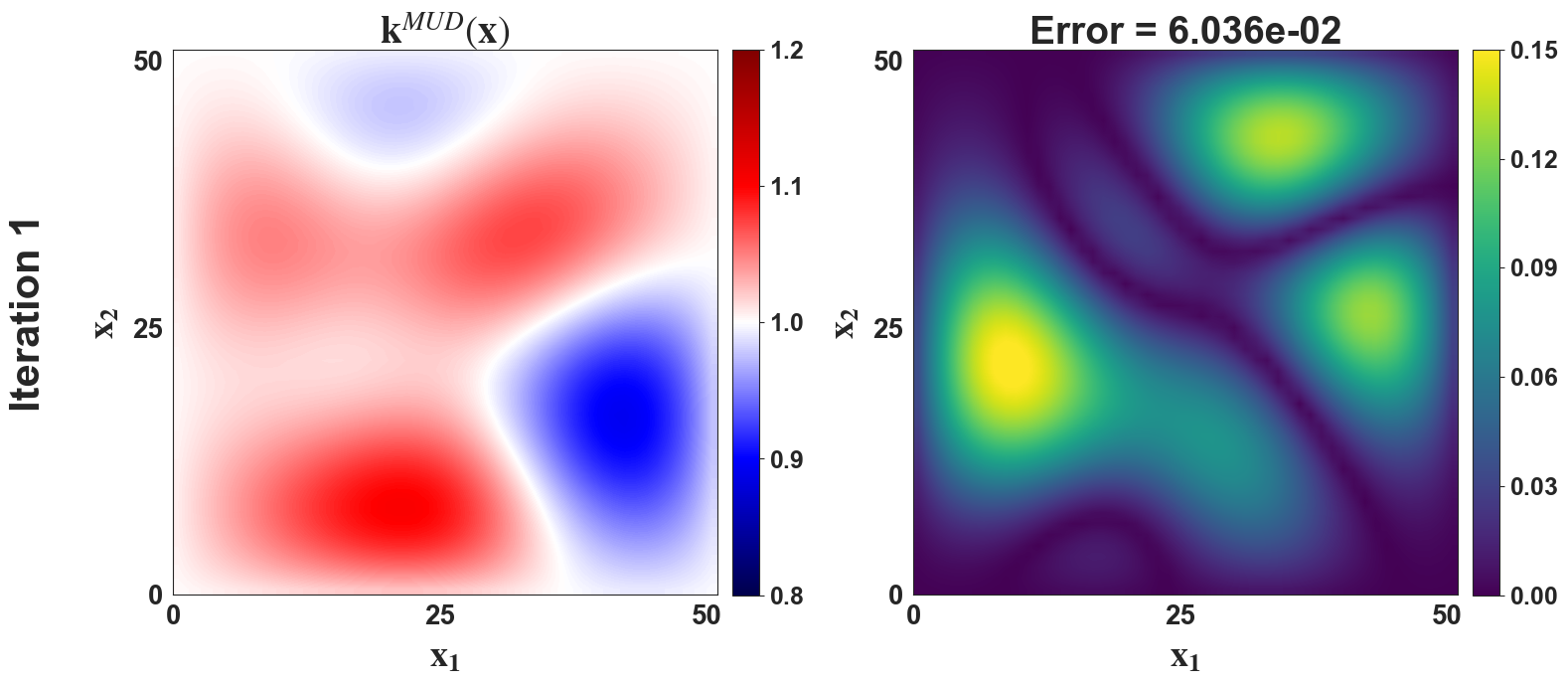}
  \end{subfigure}
  \begin{subfigure}{0.7\textwidth}
    \includegraphics[width=\linewidth]{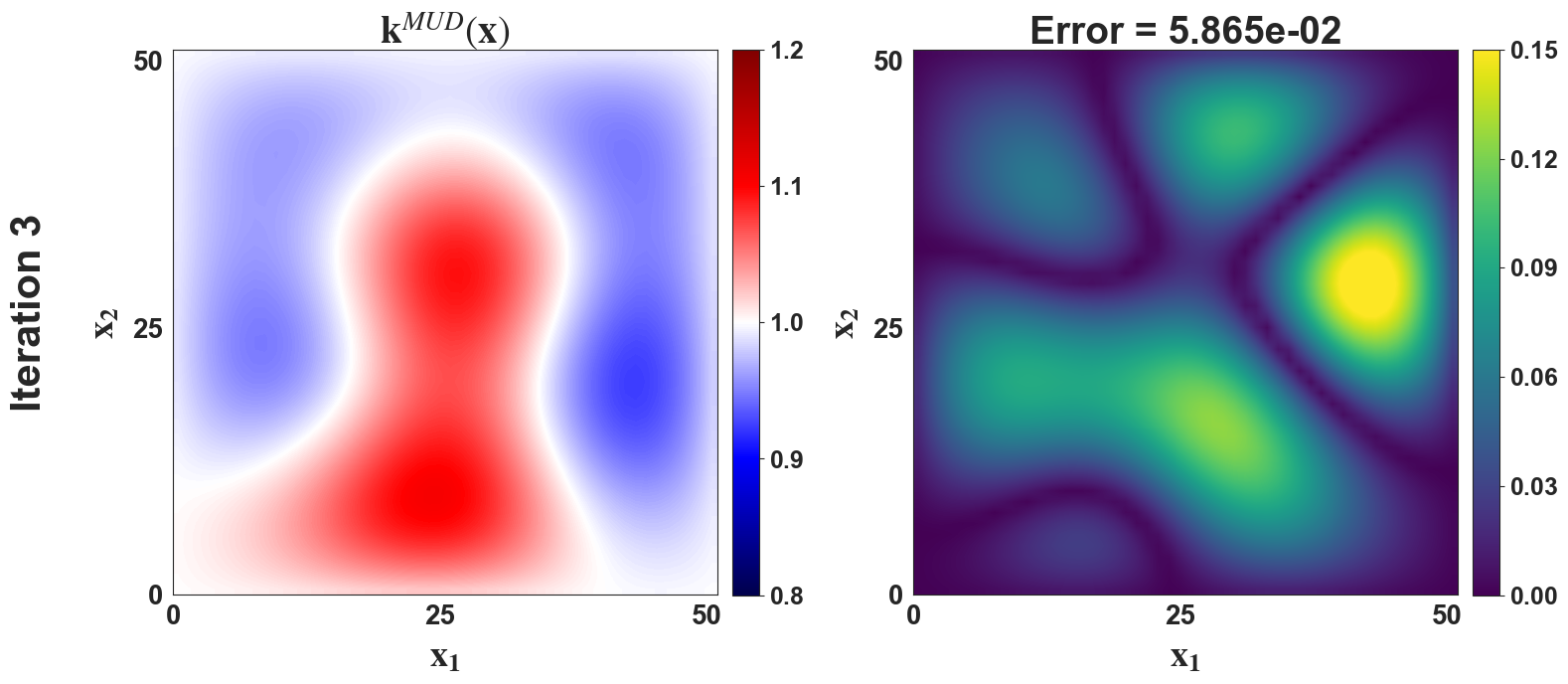}
  \end{subfigure}
   \begin{subfigure}{0.7\textwidth}
    \includegraphics[width=\linewidth]{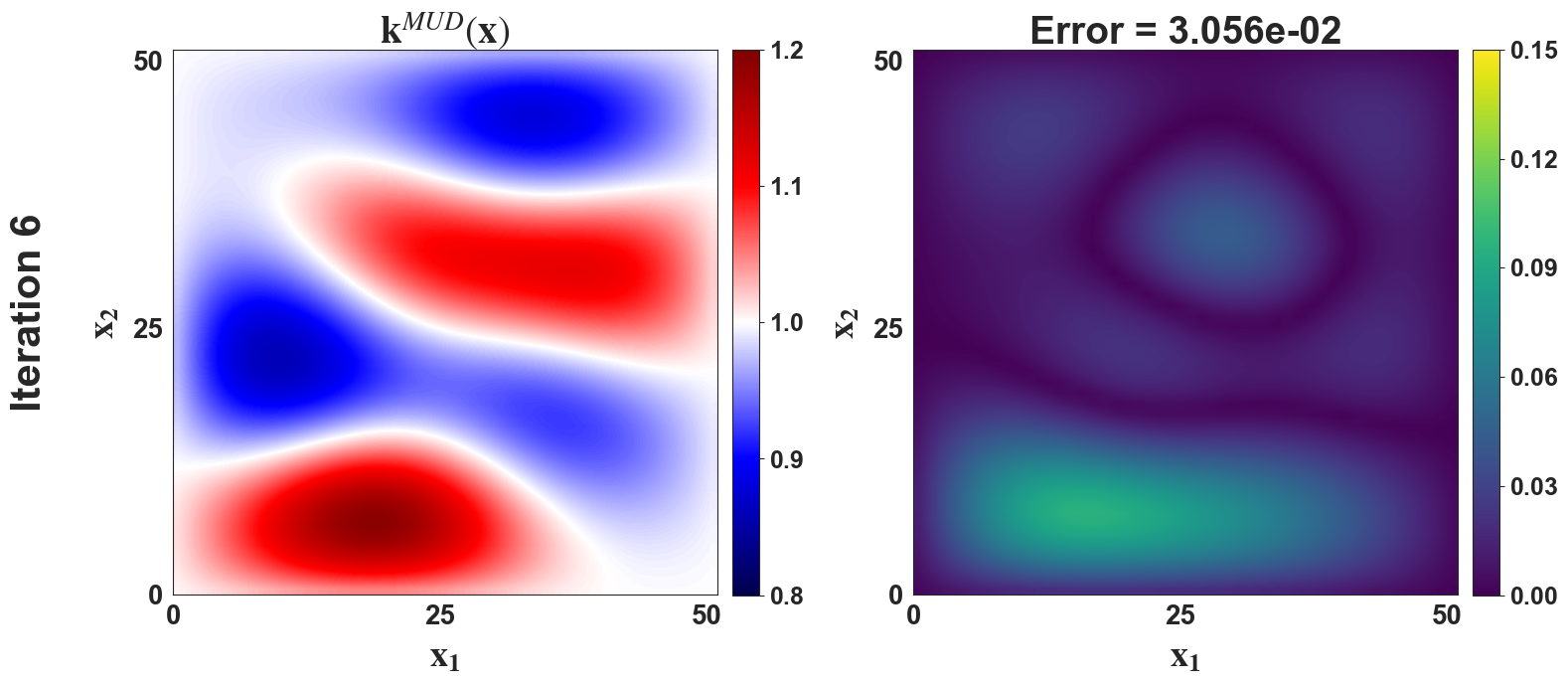}
  \end{subfigure}
  \begin{minipage}{0.7\textwidth}
        \caption{True thermal diffusivity field using 10 KL modes (top) along with estimate fields according to the $\lambda^\mathrm{MUD}$ parameter estimate (left) and the error between the true and approximate fields (right) on iterations 1, 3, and 6 ($t=0.5, 1.5, 3.0$).}\label{fig:hm_est}
  \end{minipage}
\end{figure}

Figure~\ref{fig:hm_est} shows the progression of the reconstructed thermal diffusivity fields corresponding to the $\lambda^\mathrm{MUD}$ points at different iterations.
These estimates are denoted by $\mathbf{k}^\text{MUD}(\boldsymbol{x})$ at each iteration and the true thermal diffusivity field is denoted simply as $\mathbf{k}(\boldsymbol{x})$. 
The online sequential estimation algorithm efficiently reconstructs $\mathbf{k}(\boldsymbol{x})$ as evidenced by $\mathbf{k}^\text{MUD}(\boldsymbol{x})$ containing more of the features present within $\mathbf{k}(\boldsymbol{x})$ (left plots) and the reduction in error in these MUD estimates (right plots) as the iterations increase.
We emphasize that the algorithm achieved this final estimate of the nominally 10-dimensional parameter space using a sequence of lower-dimensional QoI maps (limited to no more than three-dimensional) and a small sample size $k = 100$ than one would typically expect is required.
This is due to the aforementioned controls that allowed for efficient parameter space exploration at each iteration.
Specifically, the $\qpca$ map learns the optimal directions to perform an update to the initial density, and this updated density is subsequently utilized in the re-sampling control.

\subsection{Change Point Detection: SEIRS Model}\label{sec:seir}

\begin{figure}[h]
\centering
\begin{tikzpicture}[->,>=stealth',shorten >=1pt,auto,node distance=3cm,
  thick,main node/.style={circle,draw,
  font=\sffamily\Large\bfseries,minimum size=15mm}]

  \node[main node,fill=blue!40] (S) {S};
  \node[main node,fill=red!50] (E) [right of=S] {E};
  \node[main node,fill=green!50] (I) [right of=E] {I};
  \node[main node,fill=orange!50] (R) [right of=I] {R};

  \path[every node/.style={font=\sffamily\small,
  		fill=white,inner sep=1pt}]
    (S) edge [left=40] node[above=1mm] {\textbf{Infection}} (E)
    (S) edge [left=40] node[below=1mm] {\large$\lambda_{1}$} (E)
    (E) edge [left=30] node[above=1mm] {\textbf{Latency}} (I)
    (E) edge [left=30] node[below=1mm] {\large$\lambda_{2}$} (I)
    (I) edge [left=30] node[above=1mm] {\textbf{Recovery}} (R)
    (I) edge [left=30] node[below=1mm] {\large$\lambda_{3}$} (R)
    (R) edge [bend right=50] node[above=1.5mm] {\textbf{\textbf{Loss of Immunity}}} (S)
    (R) edge [bend right=50] node[below=1.5mm] {\large{$\lambda_{4}$}} (S);
\end{tikzpicture}
\caption{The SEIRS model with demography. Rates are $\lambda_{1}$ (contact), $\lambda_{2}$ (latency), $\lambda_{3}$ (recovery), $\lambda_{4}$ (loss of immunity). The relationship between $R_0 = \lambda_1 / \lambda_3,$ known as the basic reproduction number determines the periodicity of the model, with periodic behavior when $R_0 > 1.0$}\label{fig:seir}
\end{figure}
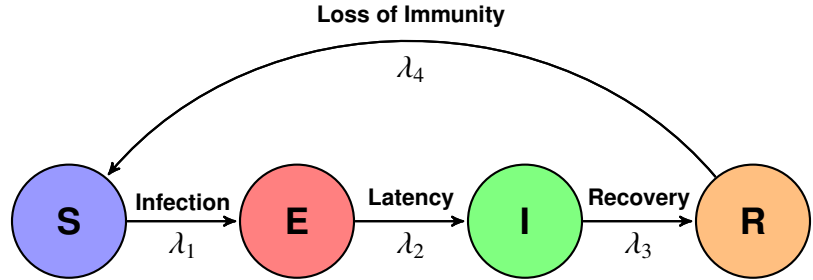

The final example considers the SEIRS model, which is a classical model used in epidemiology.
This example illustrates how the sMUD Algorithm~\ref{alg:seq_MUD} can be used for change point identification problems.
The SEIRS model is composed of four components, shown in Figure \ref{fig:seir} and modeled by the following dynamic equations relating these components:
\begin{equation}
\begin{cases}
\begin{aligned} \label{eq:seir}
\frac{\mathrm{d} S}{\mathrm{~d} t}&= \underbrace{\mu N}_{\text {birth }}-\underbrace{\lambda_{1} \frac{I S}{N}}_{\text {infection }}+\underbrace{\lambda_{4} R}_{\text {lost immunity }}\\
\frac{\mathrm{d} E}{\mathrm{~d} t}&=\underbrace{\lambda_{1} \frac{I S}{N}}_{\text {infection }}-\underbrace{\lambda_{2} E}_{\text {latency }}\\
 \frac{\mathrm{d} I}{\mathrm{~d} t}&=\underbrace{\lambda_{2} E}_{\text {latency }}-\underbrace{\lambda_{3} I}_{\text {recovery }}\\ 
\frac{\mathrm{d} R}{\mathrm{~d} t}&=\underbrace{\lambda_{3} I}_{\text {recovery }}-\underbrace{\lambda_{4} R}_{\text {lost immunity }}     
\end{aligned}
\end{cases}
\end{equation}

The proposed model represents the dynamics of an infectious disease outbreak, where the states are defined as Susceptible $(S)$, Exposed $(E)$, Infectious $(I)$, and Recovered $(R)$. 
Susceptible individuals are those who are at risk of contracting the disease, while Exposed individuals have been exposed to the virus but have not yet been infected.
Infectious individuals are those currently infected, while Recovered individuals are those who have recovered from the disease. 
The total population at any given time $\mathrm{t}$ is defined as $N = S + E + I + R = 1$, where we usually express each quantity in terms of the fraction of the total population.

The parameters within the model play a critical role in determining the outbreak dynamics,
Specifically, the infection rate $\lambda_{1}$ is a key factor that governs the speed of disease spread, while $\lambda_{2}$ denotes the incubation rate or the rate at which latent individuals become infectious, with the average incubation period being $1/\lambda_{2}$. 
Additionally, the recovery rate or mortality rate $\lambda_{3}$ determines the rate at which individuals recover from the disease or succumb to it.
If the duration of infection is $T$, then $\lambda_{3}$ can be expressed as $\lambda_{3}=1/T$. 
When $R_0 = \lambda_1 / \lambda_3 > 1$, the SEIRS model exhibits periodic behavior.
The model is subject to the initial conditions $S(0)>0$, $I(0) \geq 0$, $E(0) \geq 0$, and $R(0) \geq 0$, which reflect the fact that there must be some susceptible individuals at the start of the outbreak, and that there may be individuals who are already infected or have recovered at time $\mathrm{t}=0$.

\subsubsection{Parameter Shifts - Lockdown Policy (Shift 1) and Virus Mutation (Shift 2)}

\begin{figure}
    \centering
    \includegraphics[width=0.7\linewidth]{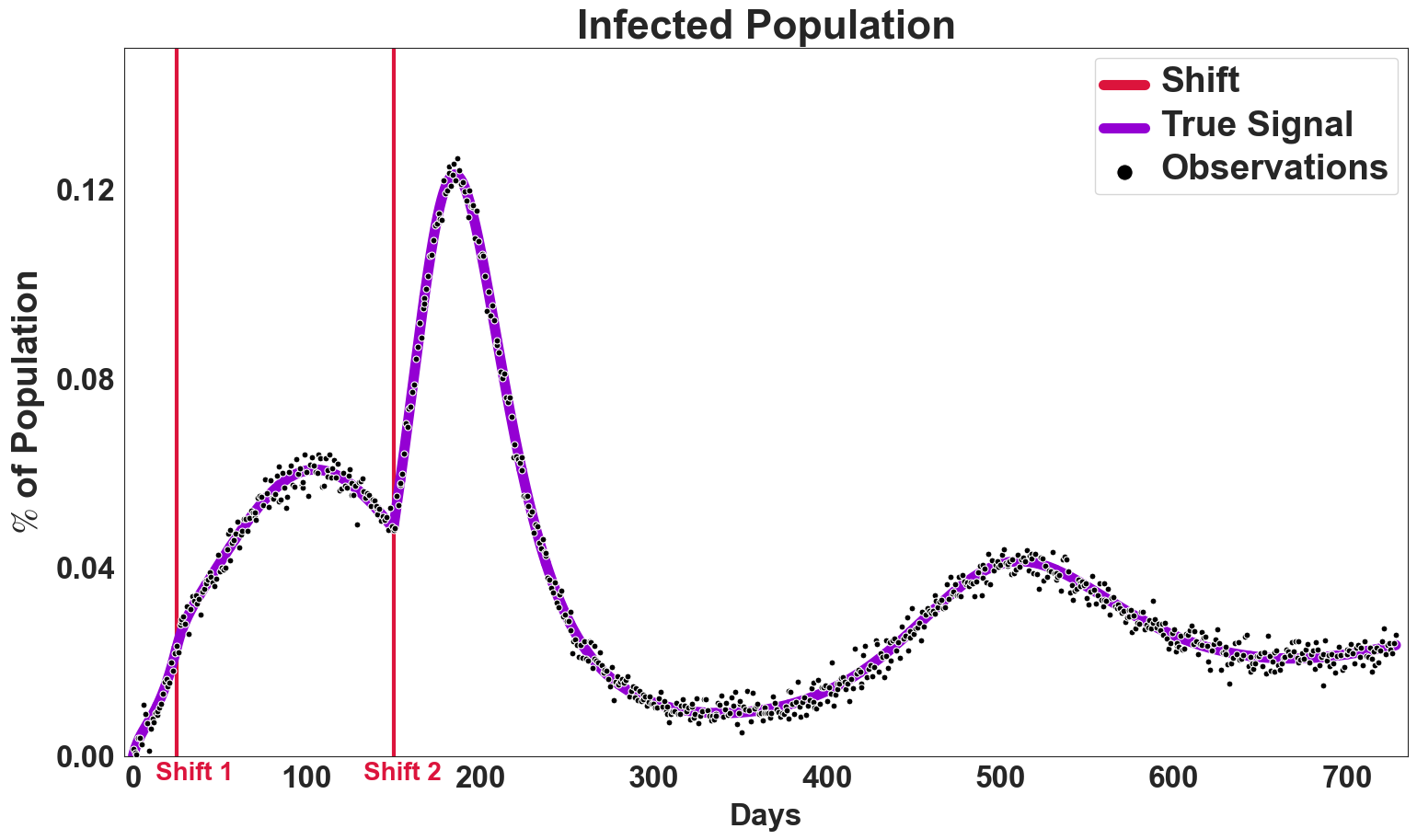}
    \caption{Infected population (purple line) over simulation of 1 year, with two parameter shifts at day 25 and day 150 indicated in veritcal red lines. We can see how the first shift corresponds to a ``flattening of the curve," the desired effect of a policy decision such as lockdowns, and the second shift corresponds to a potential ``second wave" due to a virus mutation.}
    \label{fig:seirs_infected}
\end{figure}

One of the key problems with modeling epidemics with the SEIRS model is that the parameter values rarely stay fixed over the course of the epidemic.
Policy decisions, virus mutations, and other changes in population dynamics continuously affect how the disease transmits and spreads.
Detecting when these shifts occur is critical to accurately informing policy and decision makers to best mitigate and control a breakout.
To simulate these shifting dynamics, the ``true'' parameter values in the simulation shift twice during the 1-year simulation.
First, at day 25, the transmission rate is halved, modeling the effects of a lock-down policy that may be enacted to ``flatten the curve.''
Second, at day 150, the transmission rate is increased to 1.2 times its original value, and the incubation rate of the disease if halved, modeling a virus mutation that makes the disease more infectious and with a lower latency period (exposed hosts become infected quicker).
The second shift corresponds to a ``second wave'' of an epidemic break.
See Table~\ref{tab:sim_attr} for the list of parameter values.

\begin{table}[h]
  \centering
\begin{tabular}{l *4c @{}}    \toprule
\textbf{\emph{Parameters}} & \textbf{\emph{True}}  & \textbf{\emph{Change-Point 1 (Day = 25)}}  & \textbf{\emph{Change-Point 2 (Day = 150)}}  \\\midrule
\rowcolor{blue!40}  $\boldsymbol{\lambda_{1}}$ & \textbf{3.0 / 14.0} & \textbf{0.5 /14.0} & \textbf{3.6/14.0}  \\ 
\rowcolor{red!45}  $\boldsymbol{\lambda_{2}}$ & \textbf{1.0/7.0} & \textbf{--} & \textbf{1.0 / 3.5}  \\
\rowcolor{green!45}  $\boldsymbol{\lambda_{3}}$ & \textbf{1.0/14.0} & \textbf{--} & \textbf{--} \\
\rowcolor{orange!45}  $\boldsymbol{\lambda_{4}}$ & \textbf{1.0/365.0} & \textbf{--} & \textbf{--} \\\bottomrule
 \hline
\end{tabular}
\caption{SEIRS Parameter Values with Shifts. We start with an incubation rate of one week, a recover rate $\lambda_2$ of two weeks (14 days), a loss of immunity rate $\lambda_3$ of 1 year (364 days), and an infection rate $\lambda_1$ such that $R_0 = 3$. Infection/Transmission/Contact rate ($\lambda_1$, first row) changes at both points, Latency/Incubation rate ($\lambda_2$, second row), changes only at the second change point. The other parameters, the recover rate $\lambda_3$ and loss of immunity rate $\lambda_4$ stay constant. }
\label{tab:sim_attr}
\end{table}

To frame the parameter estimation problem, we begin by generating synthetic data by initializing the ``true'' state to $S(0)=0.98$, $E(0) = 0.01$, $I(0) = 0.01$,  and $R(0) = 0$, and we use a 4th-order explicit Runge-Kutta time integrator with a time step of $\Delta t=0.1$ to propagate the dynamical system forward in time, adjusting the true parameter values as necessary according to Table~\ref{tab:sim_attr}.
Synthetic measurements of the infected population are collected daily by populating this true state with $\sigma=0.005$ levels of noise to reflect uncertainty in this data  (see Figure~\ref{fig:seirs_infected}).
Uncertainty in this data can be due to many sources, e.g., incomplete census data of either self-reported individuals or those that are sick enough to warrant hospitalization.
We emphasize that we only observe the infected population $I$, as the other states are usually not observable in a real world scenario.
The goal is to solve the parameter estimation problem using the sMUD algoithm, and track Diagnostics 1 and 3 introduced in Section~\ref{subsec:diag3} to see if we can identify the change-points a-posteriori with different values of $\epsilon_\mathrm{pred}$ and $\epsilon_{\delta - \mathrm{point}}$.

We apply the sMUD algorithm using $k = 1000$ samples, collecting batches of $\Delta t = 14$ days of data over a period of $T = 364 $ days, or one year.
We let $\initdens$ define a uniform distribution over a parameter spaced defined by intervals that are $\pm 100\%$ of the true parameters initial values (that is the first column in Table~\ref{tab:sim_attr}), and using $1\leq m \leq 3$ principle components at each iteration to then choose the solution according to Diagnostic 1 in Section~\ref{subsec:diag1} with an $\er$ estimate closest to 1.0.
Since there are parameter shifts, it is crucial to determine the appropriate circumstances for applying Controls 3 and 4.
The SEIRS model requires relatively low computational resources to solve, which naturally leads to the selection of resampling the initial distribution (Control 4).
However, for a computationally demanding forward model, such as the ADCIRC model referenced in Section~\ref{sec:adcirc}, repeatedly generating a large sample set may become impractical.
Consequently, in such instances, Control 3 emerges as the preferred method for sequentially updating the MUD estimate upon recognizing a shift in the true parameter's underlying distribution.

\begin{figure}
  \centering
  \begin{subfigure}{0.75\textwidth}
    \includegraphics[width=\linewidth]{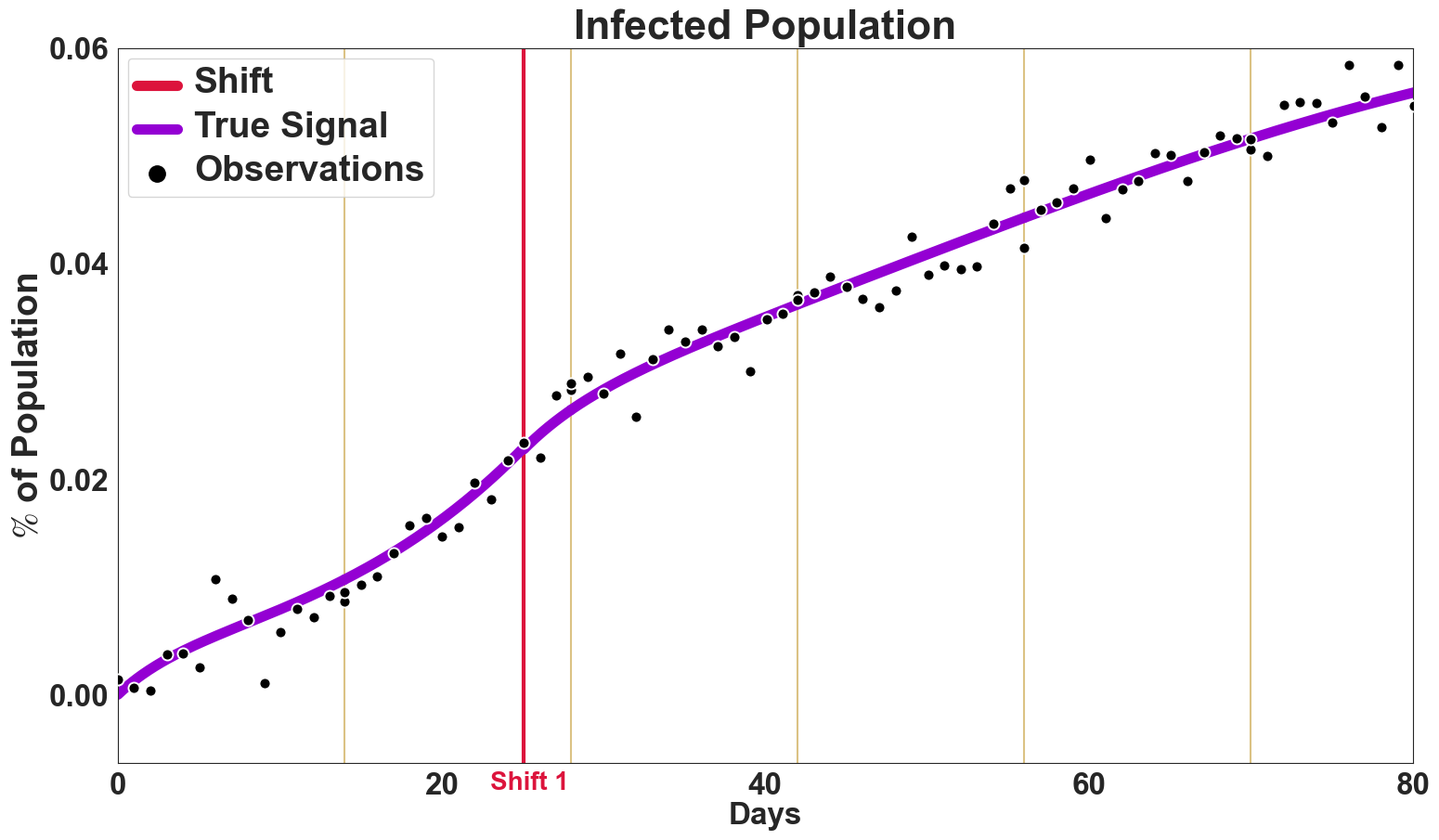}
  \end{subfigure}
  \begin{subfigure}{0.75\textwidth}
    \includegraphics[width=\linewidth]{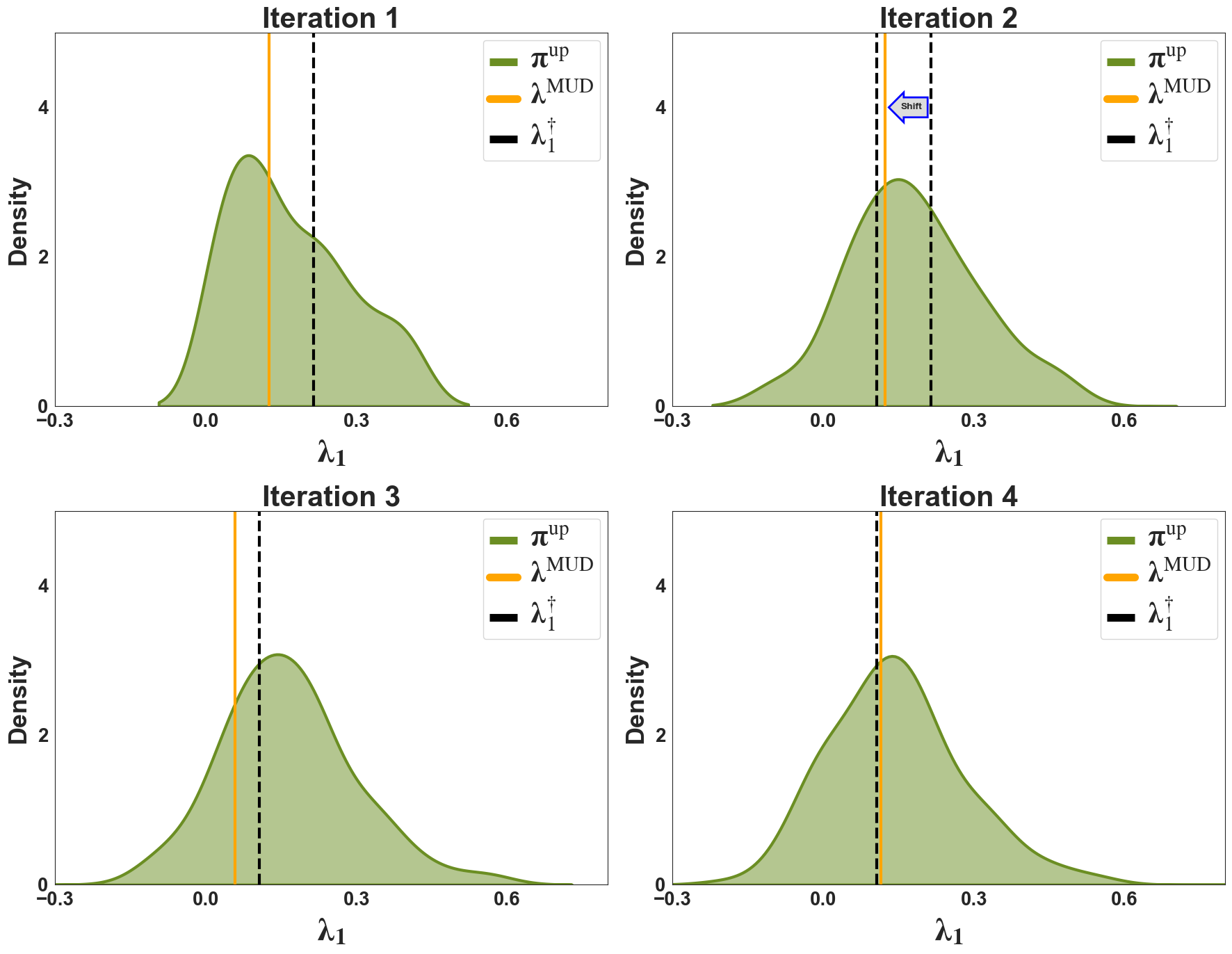}
  \end{subfigure}
  \begin{minipage}{0.75\textwidth}
    \caption{Shift 1 (Flattening of curve) $\rightarrow$ Change in transmission rate ($\lambda_1$) - (top) Infected population state over first 6 iterations, along with updated density plots for the transmission rate $\lambda_1$ for first 4 iterations (bottom). Shift happens in second iteration (indicated by arrow), where we see a decrease in the transmission rate due to a ``flattening of the curve'' as a result of a lockdown policy for example. We see how the online sequential estimation algorithm naturally shifts to peak around the new transmission rate, even though it does not really estimate the first transmission rate value in the first iteration very well due to the high amount of noise in the early data-points compared to the signal (top plot).}
    \label{fig:shift_1}
  \end{minipage}
\end{figure}

\begin{figure}  
  \centering 
  \begin{subfigure}{0.75\textwidth}
    \includegraphics[width=\linewidth]{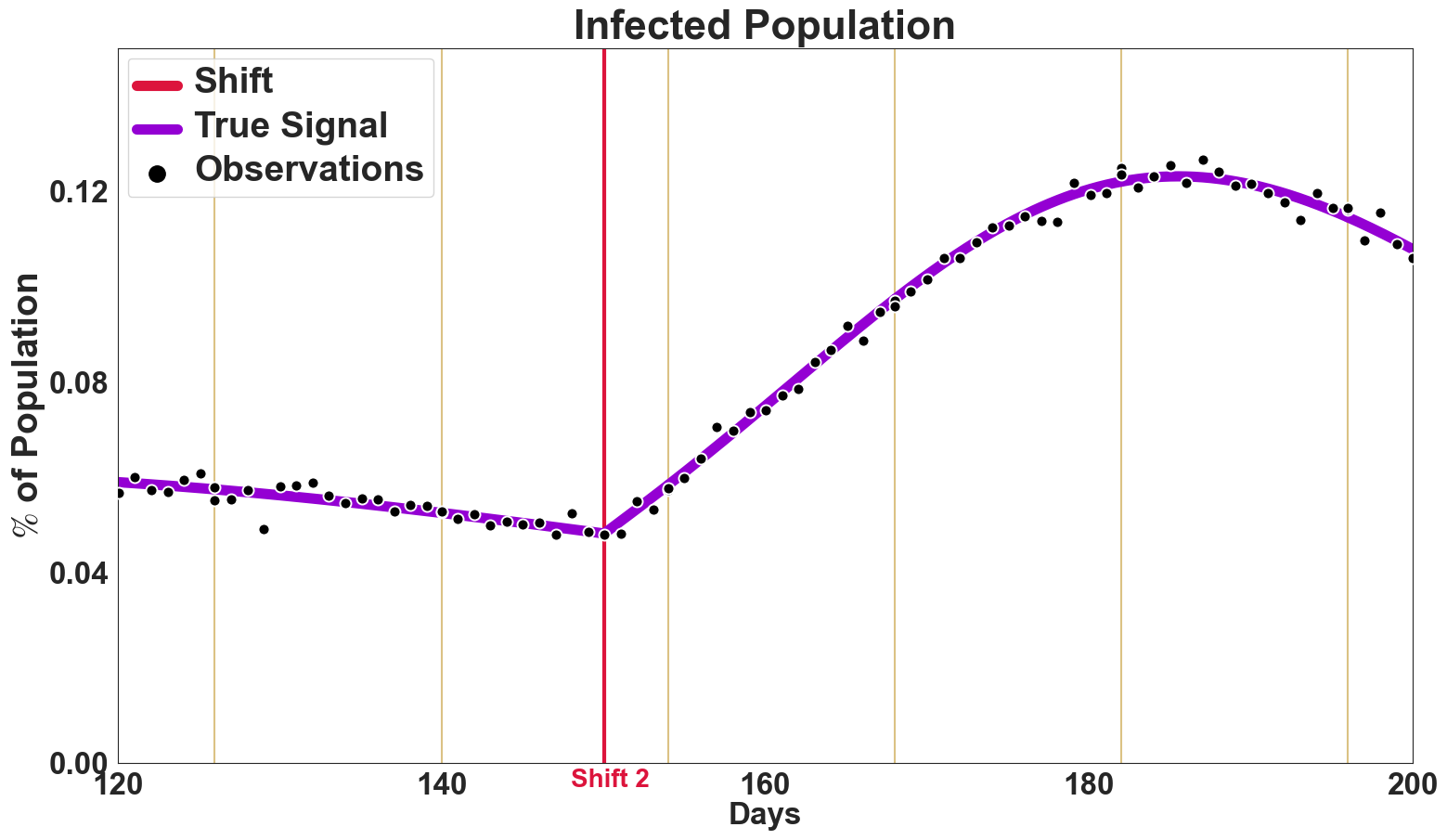}
  \end{subfigure}
  \hspace{1cm}
  \begin{subfigure}{0.75\textwidth}
    \includegraphics[width=\linewidth]{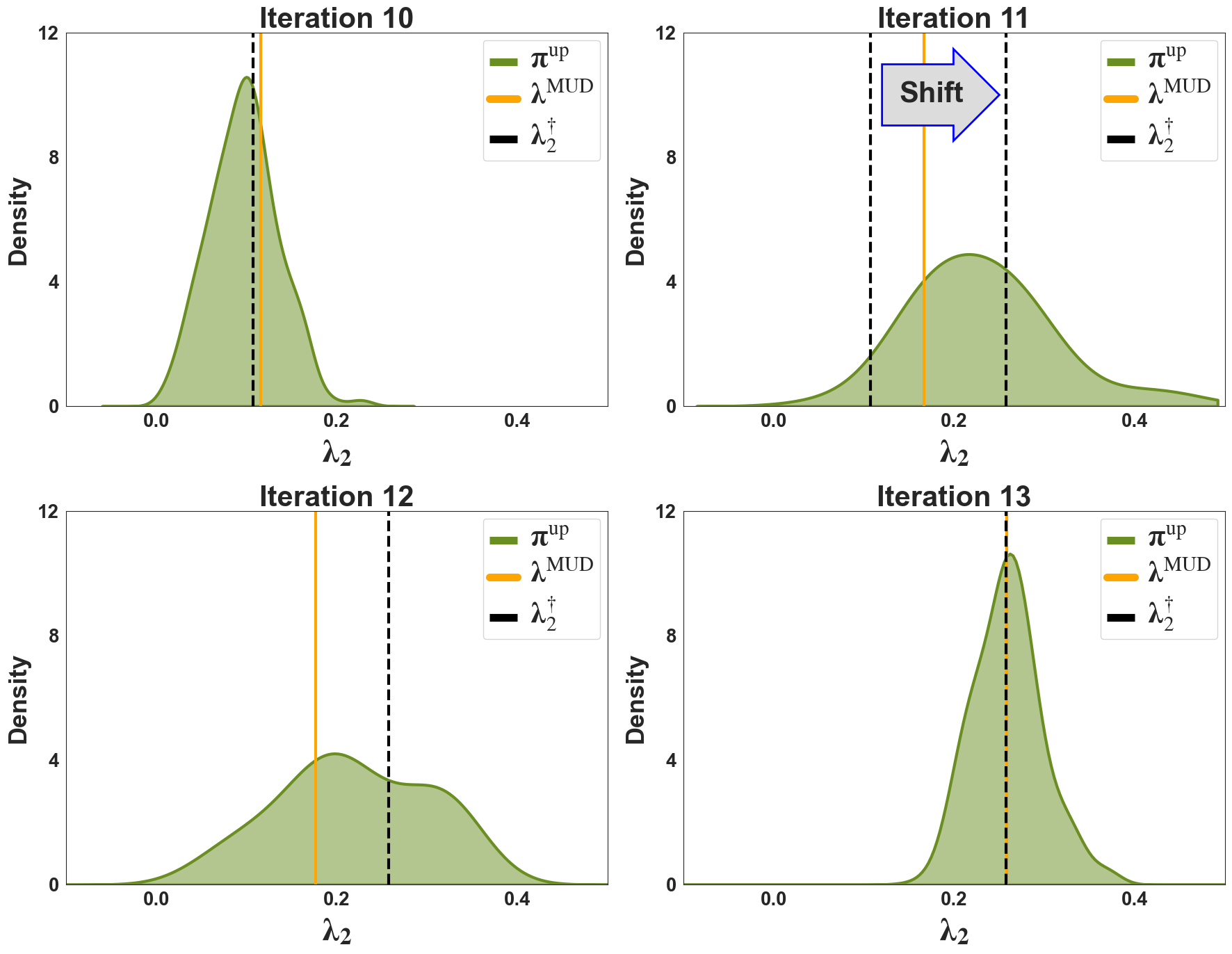}
  \end{subfigure}
  \begin{minipage}{0.75\textwidth}  
  \caption{Shift 2 (Virus Mutation) $\rightarrow$ Change in transmission and incubation rate ($\lambda_1, \lambda_2$) - (top) Infected population state from iterations 9 - 15, along with the updated density plots for the incubation rate ($\lambda_2$) at iterations 10-13. Shift occurs in the 11th iteration, at day 150, where we see in increase in the rate of incubation, or the rate at which exposed hosts become infected. Note how we are able to clearly see the shift in parameter value here most likely due to (1) the signal to noise ratio being much higher during these iterations and (2) the incubation rate being a rate directly influencing the infected state observable (as opposed to the transmission rate that affects the infected population indirectly).}  
   \label{fig:shift_2}
  \end{minipage}
\end{figure}

Figures~\ref{fig:shift_1} and ~\ref{fig:shift_2} show the progression of updated densities for the iterations around the first and second shift, respectively.
Note how in both cases, the updated distribution naturally centers and concentrates around the shifted values as iterations progress.
In the first shift case, when the transmission rate is being halved, we see that there is difficulty in estimating $\lambda_1$ prior to the shift for two reasons: (1) this shift occurs relatively soon after the start of the simulation, and (2) the signal-to-noise ratio during these initial iterations is relatively high (Figure~\ref{fig:shift_1}).
For the second shift, where the infection and incubation/latency rates both change, we see a very distinct shift in the updated distribution from centered around the prior true parameter value to the curren true parameter value around the iterations where the shift occurs (note the plot only shows the latency rate $\lambda_2$ for brevity). 

\begin{figure}
  \centering
  \begin{subfigure}{0.8\textwidth}
    \includegraphics[width=\linewidth]{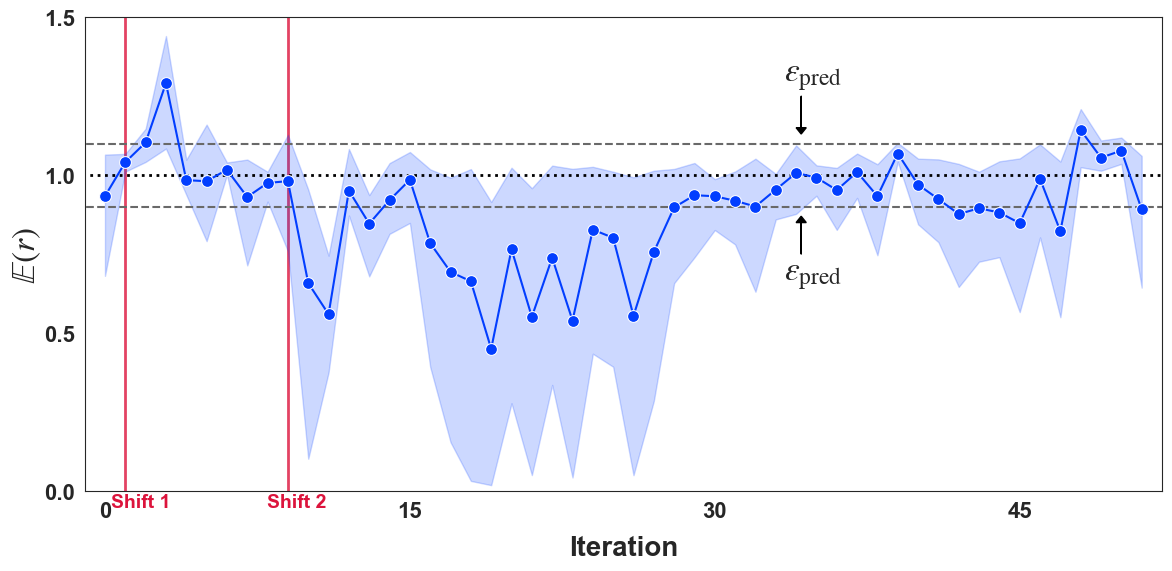}
    \label{fig:seirs_er_kl}
  \end{subfigure}
  \hspace{1cm}
  \begin{subfigure}{0.8\textwidth}
    \includegraphics[width=\linewidth]{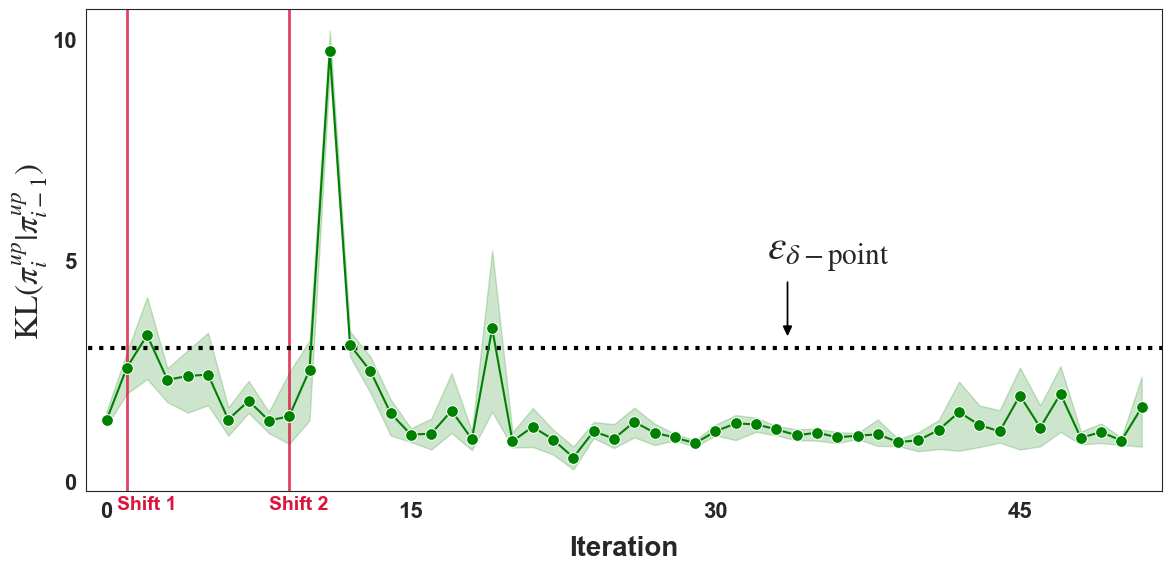}
    \label{fig:seirs_kl}
  \end{subfigure}
  \begin{minipage}{0.75\textwidth}
  \caption{Change point metrics over time - (top) $\mathbb{E}(r)$ over each iteration, which should be $\epsilon_\mathrm{pred} = 0.1$ away from $1.0$ (indicated in dashed horizontal lines around the desired black dashed line). (bottom) $KL_\mathrm{DCI}$, which should not exceed $\epsilon_{\delta-point}$ on any given iteration. The iterations in which the two shifts occur are indicated in solid red lines. Note the slight lag in when the shifts occurs and the change point conditions are violated.}  \label{fig:seirs_cpi}
  \end{minipage}
\end{figure}

As for determining when the shifts occur, Figure~\ref{fig:seirs_cpi} shows the progression of $\er$ and the information gain $\mathrm{KL}(\updens^{(i)}|\updens^{(i-1)})$.
We note that the spread in values at each point in the graph corresponds to the variability in all the combinations of offline sequential estimation attempted.
Taking $\epsilon_\mathrm{pred} = 0.1$ and $\epskl = 3$, we see that the two shifts line up well with the change point criteria.
We also note that there is an expected lag of about 1 to 2 iterations for detecting the change points, as they occur in the middle of estimation iterations and the system requires some time for state variables to exhibit changes due to the changing parameter values.
Furthermore, we note that we misidentify a change point towards iteration 19, illustrating the trade-off of setting the information gain threshold $\epskl$ too low. 
Had we set it higher, we risk missing the first shift since it was more subtle in terms of the changes in the observed dynamics compared to those observed in the second shift where a relatively large information gain spike is observed. 
Overall, we see the importance of using the combination of the information gain with the expected ratio to inform change points as there are many iterations (15-30) where $\er$ is well outside the specified threshold. 
These iterations correspond to times in the epidemic simulation (days 210 - 420) where the infection rates are the lowest and almost constant.
This is where we expect poorer estimates of $\er$ values, especially when trying to use a higher number of principal components to define the QoI map.
Specifically, it is here that the system states are not as sensitive to small perturbations in the parameters, which implies it is more difficult to estimate distributions associated with a high-dimensional QoI map.




\section{Conclusion and Future Work}\label{sec:conclusions}

We presented a novel algorithm with various mathematically justified diagnostics serving to control a sequential data-consistent approach to parameter estimation for dynamical systems.
Numerical examples demonstrated the applicability of this approach in various scenarios including an offline vs online scenario and to detect and estimate change points for parameters that drift or shift suddenly in time.
We specifically applied the algorithm to three distinct problems.
In the first, we showed how offline sequential estimation can accurately estimate a set of wind-drag parameters for a high-fidelity storm-surge model.
The results demonstrate that comparable results are obtained using an iterative update with a scalar-valued QoI map as opposed to a non-iterative update that utilizes all available data to construct a vector-value QoI map. 
In the second example, we illustrate how sequential estimation can solve a higher-dimensional parameter estimation problem using noisy spatiotemporal data to estimate the thermal diffusivity field in a heat conductivity problem where the field is parameterized using a KL-expansion with 10 KL modes.
This example also demonstrated how re-sampling at iterations can provide efficient exploration of a high-dimensional parameter field when iterating over a sequence of lower-dimensional QoI maps.
Finally, we applied the sequential algorithm to an epidemiological model involving two distinct change-points related to changes in policy and a virus mutation.
This example demonstrated the ability of the algorithm to respond to change-points by adjusting parameter estimates and the associated uncertainty in these estimates. 
Moreover, this example illustrated that we can accurately detect change-points with the diagnostics utilized in the algorithm. 

The algorithms presented in this work as well as the open source python package \href{https://github.com/UT-CHG/pyDCI/tree/master}{pyDCI} that accompanies this work (see ~\ref{sec:software}) can serve as a practical guide for practitioners to learn and start applying data-consistent inversion to various applications.
Future research will focus on the theoretical analysis and performance of the sequential algorithm presented, both in terms of convergence criteria and error bounds.
Another topic for future work is to connect the choice of hyper-parameters in the algorithms presented to concepts in the field of Optimal Experimental Design (OED).
Furthermore, there are numerous connections and comparison studies to be made with the more traditional data assimilation techniques and change-point detection algorithms. 


\section{Acknowledgments}\label{sec:acknowledge}

T.~Butler's work is supported by the National Science Foundation under Grant No.~DMS-2208460.
T.~Butler's work is also supported by NSF IR/D program, while working at National Science Foundation. 
C.~Dawson’s, R.~Spence's and C.~del-Castillo-Negrete’s work is supported in part by the National Science Foundation No. DMS-2208461. 
However, any opinion, finding, conclusions, or recommendations expressed in this material are those of the authors and do not necessarily reflect the views of the National Science Foundation.
The authors would also like to acknowledge the usage of the compute resources at the Texas Advanced Computing Center (TACC), and in particular to the DesignSafe project infrastructure~\cite{rathje_designsafe_2017} for hosting the ADCIRC data-set used in this research.
Finally the authors would like to thank Dr. Michael Pilosov for the invaluable insights he provided in the early discussions of a sequential parameter estimation framework.


\bibliographystyle{elsarticle-num-names}
\bibliography{main.bib}

\appendix

\section{Software Contributions}\label{sec:software}

All work presented is available via the open source \href{http://www.github.com/UT-CHG/pyDCI}{github.com/UT-CHG/pyDCI} python package and published to the PyPi Pyhon Package Registry under the name pyDCI.
The library provides the following features:

\begin{itemize}
    \item A set of python classes that build off of one another to solve Data-Consistent Inversion problems, with a focus on sequential parameter estimation.
    \item Logging and error catching, with reasonable exceptions and errors raised when diagnostics such as the predictability assumption or parameter drift are detected.
    \item An extensive set of notebooks that expand on the examples presented in this work.
\end{itemize}

Running `pip install pyDCI' will install the pyDCI package and its dependencies, and expose a Command Line Interface (CLI) to run the main examples presented in this work.

\end{document}